\documentclass[fleqn,usenatbib,useAMS]{mnras}

\usepackage{lmodern}
\usepackage{graphicx}	
\usepackage{amsmath}	
\usepackage{amssymb}	
\usepackage{multicol}        
\usepackage{bm}		
\usepackage{pdflscape}	
\usepackage{physics}
\usepackage{xspace}
\usepackage{color}
\usepackage{tikz}
\usepackage{orcidlink}
\usepackage{xcolor}
\usepackage{url}
\usepackage{hyperref}

\usetikzlibrary {shapes.geometric} 

\definecolor{newblue}{RGB}{37, 150, 190}
\definecolor{newyellow}{RGB}{219, 172, 90}
\definecolor{newred}{RGB}{209, 23, 23}

\newcommand{\comet}{\mathrm{comet}}
\newcommand{\crit}{\mathrm{crit}}

\newcommand{\Jup}{\mathrm{Jup}}

\usepackage[T1]{fontenc}
\usepackage{ae,aecompl}

\usepackage{newtxtext,newtxmath}

\title[Polluting White Dwarfs with Oort Cloud Comets]{Polluting White Dwarfs with Oort Cloud Comets}

\author[D. Pham, H. Rein]{
Dang Pham$^{1}$\thanks{E-mail: dang.pham@astro.utoronto.ca (DP)}\orcidlink{0000-0002-0924-8403},
Hanno Rein$^{1,2}$ \orcidlink{0000-0003-1927-731X}\\
$^1$ Department of Astronomy and Astrophysics, University of Toronto, Toronto, Ontario, M5S 3H4, Canada\\
$^2$ Department of Physical and Environmental Sciences, University of Toronto at Scarborough, Toronto, Ontario M1C 1A4, Canada
}

\date{Accepted 2024 April 05. Received 2024 April 05; in original form 2024 March 06}

\pubyear{2024}

\begin{document}
\label{firstpage}
\pagerange{\pageref{firstpage}--\pageref{lastpage}}
\maketitle

\begin{abstract}
Observations point to old white dwarfs (WDs) accreting metals at a relatively constant rate over 8~Gyrs. Exo-Oort clouds around WDs have been proposed as potential reservoirs of materials, with galactic tide as a mechanism to deliver distant comets to the WD's Roche limit.
In this work, we characterise the dynamics of comets around a WD with a companion having semi-major axes on the orders of 10 - 100 AU. 
We develop simulation techniques capable of integrating a large number ($10^8$) of objects over a 1~Gyr timescale. 
Our simulations include galactic tide and are capable of resolving close-interactions with a massive companion.
Through simulations, we study the accretion rate of exo-Oort cloud comets into a WD's Roche limit.
We also characterise the dynamics of precession and scattering induced on a comet by a massive companion.
We find that 
(i) WD pollution by an exo-Oort cloud can be sustained over a Gyr timescale, 
(ii) an exo-Oort cloud with structure like our own Solar System's is capable of delivering materials into an isolated WD with pollution rate $\sim 10^8 \mathrm{~g~s^{-1}}$,
(iii) adding a planetary-mass companion reduces the pollution rate to $\sim 10^7 \mathrm{~g~s^{-1}}$,
and (iv) if the companion is stellar-mass, with $M_p \gtrsim 0.1 M_\odot$, the pollution rate reduces to $\sim 3 \times 10^5 \mathrm{~g~s^{-1}}$ due to a combination of precession induced on a comet by the companion, a strong scattering barrier, and low-likelihood of direct collisions of comets with the companion.

\end{abstract}




\section{Introduction}
Between 25 to 50 percent of white dwarf (WD) atmospheres observed are polluted with heavy metals \citep[e.g.][]{Zuckerman2003, Zuckerman2010, Koester2014, Wilson2019}.
It is generally believed that the heavy metals of polluted WDs come from its evolved planetary system, such as exomoons, exoplanets or exo-asteroid belts, through a variety of mechanisms which induce destabilisation in these sources \citep[e.g.][]{Debes2002, Debes2012, Mustill2014, Smallwood2018, Maldonado2020, Trierweiler2022, OConnor2022}.
The bodies must be delivered to the WD Roche radius at $\sim 1 R_\odot$ to be tidally disrupted for pollution \citep[e.g.][]{VerasHeng2020}.

Observations of old polluted WDs (WDs with cooling age older than 1 Gyr) found accretion rates ranging five orders of magnitude between $5 \times 10^5$ to $10^{10}  \mathrm{~g~s^{-1}}$; the mean pollution rate is around $10^{7} \mathrm{~g~s^{-1}}$ with about a 1 dex spread  \citep[e.g.][]{BlouinXu2022, Johnson2022, Cunningham2022}.
The current minimum detection limit is about a few $10^5 \mathrm{~g~s^{-1}}$ \citep{Koester2014}. 
\cite{BlouinXu2022} also find that the WD accretion rate decreases by no more than one order of magnitude over 8 Gyr.
Thus, their observational findings require a reservoir and mechanism that can deliver materials in the WD's Roche radius over such long timescales.

The chemical composition of accreted materials can be analysed to study the original reservoir of the accreted bodies.
Until now, a few dozen WDs have been followed up spectroscopically to measure their element abundances \citep{Jura2014}.
The majority of polluted WDs exhibit compositions resembling the bulk Earth. This suggests that the sources of materials polluted WDs must be rocky in nature \citep{Jura2006, Jura2012, Xu2017, Doyle2019, Trierweiler2023}.
There are observations of polluted WD atmospheres with volatiles, although they are much more rare \citep{Farihi2013, Klein2021, Doyle2021}.
At the present, there is one observation of a polluted WD with an icy body composition \citep{Xu2017}.
Because the majority of observations point to rocky materials as a source, the Oort cloud has often been ruled out as a reservoir for WD pollution. 

The Oort cloud is a byproduct of Solar System formation, where planetesimals are either ejected or kicked into high semi-major axes via interactions with a giant planet \citep{Vokrouhlicky2019, Kaib2022}.
For objects kicked into higher semi-major axes, they can be circularised by galactic tide or stellar flybys \citep{Duncan1987, Hahn1999, Higuchi2015, Vokrouhlicky2019}.
These objects form the Solar System's Oort cloud, which is a structure containing $10^{11} - 10^{12}$ objects with a total mass of $\sim 2 M_\oplus$ with semi-major axes ranging between $3~000$ AU to $100~000$ AU \citep[e.g.][]{Weissman1983, Boe2019}.

Since the Solar System's Oort cloud is a byproduct of interactions of planetesimals with planets, it is not unreasonable to expect Oort clouds to exist around other main-sequence and even WD planetary systems. 
As a result, several studies have  investigated if the Oort cloud is a suitable reservoir for pollution.
They often employ a mix of numerical and analytic methods to study the pollution rate of Oort cloud comets.
Mechanisms considered consist of galactic tidal effects, stellar flybys, WD natal kicks and stellar mass loss during post-main-sequence evolution \citep{Alcock1986, Parriott1998, Veras2014, Stone2015, Caiazzo2017}.
Simulations of comets into a $1 R_\odot$ tidal radius in existing literature we are aware of do not resolve pollution rate over time due to being resolution limited in the number of simulated comets.
Most recently, \cite{OConnor2023} perform an extensive analytic and numerical studies on the effects of galactic tide, planetary perturbations, WD natal kicks and stellar mass loss on the pollution rate of WDs with exo-Oort clouds as a source of materials.
They find that combining these effects together point to an accretion rate of a few $10^5 \mathrm{~g~s^{-1}}$, which is just at or slightly below the detection limit.
They argue that this is potentially why we do not observe many volatiles-rich polluted WD atmospheres.

In this article, we contribute to this existing line of investigation by answering the following question through numerical simulations: 
Can an Oort cloud similar to the one we have in the present-day Solar System pollute a WD over a Gyr timescale? We also investigate this question in cases where the WD has a planetary-mass or a stellar-mass companion. Since we currently have no knowledge of extrasolar Oort clouds, we simply assume throughout this article that other Oort clouds have the total number of objects and total mass like the current Solar System Oort cloud. 
We also study how various Oort cloud powerlaw density profiles affect pollution.
All of our results can be easily scaled to another Oort cloud with different mass, number of objects, and density profile.

We start by presenting the analytic theory of WD pollution through galactic tide and WD companion (planet or star). Specifically, in Section \ref{sec:analytic_theory_gt}, we summarise the analytic vertical galactic tide model \citep{HT1986}, the loss cone theory, leading to the expected analytic pollution rate as predicted by \cite{OConnor2023}. In Section~\ref{sec:analytic_theory_companion}, we study additional dynamics comets experience when there is a companion. Specifically, we analyse the dynamics of galactic tide together with precession and scattering induced by a companion. We also summarise the loss cone shielding model proposed by \cite{OConnor2023} leading to a prediction of the WD pollution rate in the presence of a companion in that framework.

In Section \ref{sec:simulation_method}, we describe our simulation methodology which is capable of integrating a large number of comets ($10^8$ comets) over a long time (1 Gyr), initial conditions, and boundary conditions.

In Section \ref{sec:wd_pollution}, we present pollution rate over various Oort cloud structures, in the presence of galactic tide only. Then, we present the pollution rate in the presence of a stellar companion, and in the presence of a planetary companion. Finally, we show the pollution rate over a 1 Gyr timescale. In Section \ref{sec:wd_pollution}, we also compare and discuss our results with analytic expectations from \cite{OConnor2023}.

In Section \ref{sec:discussion}, we discuss advantages and major concerns of using an Oort cloud as a potential reservoir for WD pollution, such as if an Oort cloud can survive post-main-sequence evolution and that the majority of observed WDs are volatiles-poor. We also discuss our results in contexts of observations of close-in binaries, wide binaries. In Section \ref{sec:conclusion}, we summarise our findings.

\subsection{Notations}

In this work, we denote $M_*$ as the WD mass. We denote the orbital elements of a comet as: $a$ for the semi-major axis, $e$ for eccentricity, $I$ for inclination, $\omega$ for argument of pericentre, $\Omega$ for longitude of ascending node, and $l$ for mean anomaly. Orbital elements $\omega$ and $I$ are measured relative to galactic plane. Some other quantities used to describe the comet orbits are: $q$ for the pericentre distance, $Q$ for the apocentre distance, $P$ for the orbital period. The comet is assumed to be a mass-less test particle. Orbital elements with the subscript $p$ denote that they are the orbital elements for the WD companion (e.g. $M_p$ is the companion mass). We also regularly use the following set of Delaunay action-angle variables (quantities $\Lambda, L, L_z$ are actions in units of specific angular momentum, and $l, \omega, \Omega$ are angles):
\begin{align}
    \Lambda_{~} &~= \sqrt{G M_* a}          &l      \nonumber\\
    L_{~}       &~= \Lambda \sqrt{1-e^2}  &\omega \nonumber\\
    L_z     &~= L \cos I              &\Omega
\end{align}
$\Lambda$ is referred to as the circular angular momentum, $L$ as the angular momentum, and $L_z$ as the $z$ component of the angular momentum.

We use the terms `exo-Oort cloud' and `Oort cloud' interchangeably.
An exo-Oort cloud is presumed to start at the inner semi-major axis edge, $a_1$, ends at $a_2$, follows a powerlaw number density profile $n(a) \propto a^{-\gamma}$, has $N_\mathrm{Oort}$ comets and a total cloud mass $M_\mathrm{Oort}$.

When referring to our own Solar System's Oort cloud, we state `Solar System Oort cloud' explicitly.
In the Solar System, estimates for $N_\mathrm{Oort}$ typically range between $10^{11} - 10^{12}$ comets \citep[e.g.][]{Francis2005, Boe2019}.
We assume $N_\mathrm{Oort} = 10^{11}$ in this article.
It is estimated that $M_\mathrm{Oort} \sim 2 M_\oplus$ for the Solar System Oort cloud, which we will use as our fiducial value.
In addition, numerical simulations show that the Solar System Oort cloud has a powerlaw exponent $\gamma = 3.5$ \citep[e.g.][]{Duncan1987, Higuchi2015, Vokrouhlicky2019}.
Note that the total mass of the Solar System Oort cloud is also quite uncertain from simulations and observations of incoming long-period comets \citep[e.g.][]{Weissman1983,Boe2019}.

A `Solar System-like Oort cloud' is an exo-Oort cloud with $a_1, a_2, \gamma, N_\mathrm{Oort}, M_\mathrm{Oort}$ exactly like our own Solar System's Oort cloud.

\section{Analytic Theory: Galactic Tide} \label{sec:analytic_theory_gt}

\subsection{Vertical Tide Model}

\cite{HT1986} study how the galactic tide affects Oort cloud comets through an analytic model.
There are three main assumptions used in their model.
First, the star-comet system moves in a circular orbit around the galaxy.
Second, the most important component of the galactic tidal force is in the $z$ direction. 
$z$ is defined as the direction orthogonal to the galactic midplane\footnote{Because of the way $z$ is defined in our coordinate system, the inclination $I$ and argument of pericentre $\omega$ are measured relative to the $x-y$ plane parallel to the galactic midplane, instead of the usual measurement relative to the ecliptic \citep[c.f.][]{HT1986,Tremaine2023}.
All $I$ and $\omega$ used throughout this article follow this convention.}, with the midplane defined at $z=0$.
In the Solar System, the second assumption is valid since galactic tidal terms in the $x$ and $y$ components are about one order of magnitude lower than the $z$ term.
Since observations of polluted WDs are typically within the Solar Neighbourhood, the second assumption is likewise not unreasonable to apply in nearby WD planetary systems. 
Third, the galactic tidal potential experienced by the comet is approximated as the potential inside a homogeneous slab with constant density $\rho_0$.
$\rho_0$ is the averaged background density of gas and stars in the galaxy. 
As a star system orbits around the galaxy, it oscillates up and down the galactic midplane. 
As a result, $\rho_0$ also varies over time.
Following previous work \citep[e.g.][]{HT1986, OConnor2023, Tremaine2023}, we adopt an averaged fiducial value of $\rho_0 = 0.1 M_\odot ~\mathrm{pc^{-3}}$ \citep{Holmberg2000, McKee2015}. 

The galactic potential with these assumptions can be written as \citep{HT1986}:
\begin{equation}
    \Phi_\mathrm{GT} = 2 \pi G \rho_0 z^2.
\end{equation}
This potential can be averaged over the orbit of the comet and then written in Delaunay elements as:
\begin{equation}
    \langle \Phi_\mathrm{GT} \rangle = \frac{\pi \rho}{G M_*^2} \left(\frac{\Lambda}{L}\right)^2 (L^2 - L_z^2) (L^2 + 5(\Lambda^2 - L^2)\sin^2 \omega)
\end{equation}
from which we find the secular (orbit-averaged) equations of motion by applying Hamilton's equations:
\begin{equation}\label{eqn:Ldot_GT}
    \left\langle \frac{\dd L}{\dd t} \right\rangle_\mathrm{GT} = -\frac{5 \pi G \rho}{(G M_*)^2} \left(\frac{\Lambda}{L}\right)^2(L^2 - L_z^2) (\Lambda^2 - L^2) \sin(2 \omega)
\end{equation}

\begin{align}\label{eqn:omegadot_GT}
    \left\langle \frac{\dd\omega}{\dd t}\right\rangle_\mathrm{GT} = \frac{2\pi G \rho}{(G M_*)^2}\left(\frac{\Lambda}{L}\right)^2 &\left( \frac{L_z^2}{L} (L^2 + 5 (\Lambda^2 - L^2) \sin^2 \omega \right. +\nonumber \\
    &~\left.~ (L^2 - L_z^2) (L - 5 L \sin^2 \omega) \right)
\end{align}

\begin{align}\label{eqn:Omegadot_GT}
    \left\langle \frac{\dd\Omega}{\dd t}\right\rangle_\mathrm{GT} = \frac{2\pi G \rho}{(G M_*)^2} \left(\frac{\Lambda}{L}\right)^2 L_z ~\left(L^2 (5 \sin^2\omega -1) - 5 \Lambda^2 \sin^2\omega \right).
\end{align}

$\dot{L}_\mathrm{GT}$ and $\dot{\omega}_\mathrm{GT}$ are coupled differential equations. 
The phase space evolution described by these coupled equations is studied in detail in \cite{HT1986, Tremaine2023}. 
Briefly, they show that through these equations of motion, the comet can oscillate in the $(L, \omega)$ phase space due to galactic tide. 
There are two conserved Delaunay elements. $\Lambda$ is conserved because the mean anomaly is not in $\langle \Phi_\mathrm{GT} \rangle$.
Thus, the comet's semi-major axis is conserved under the galactic tidal effect. Similarly, $L_z$ is conserved because $\Omega$ is not in $\langle \Phi_\mathrm{GT} \rangle$.
Since $L_z$ is conserved but $L$ is not conserved, this implies that galactic tide can excite the comet to very high eccentricity by exchanging eccentricity with inclination.
This inclination-eccentricity exchange is periodic and is analogous to the von Zeipel-Lidov-Kozai mechanism \citep{vonZeipel1910, Lidov1962, Kozai1962}.

\subsection{Loss Cone Theory and Comet Injection Rate}

\begin{figure}
    \centering
    \includegraphics[width=\the\columnwidth]{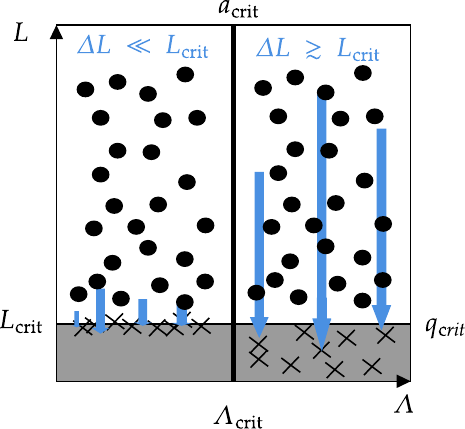}
    \caption{Diagram illustrating the loss cone theory by \protect\cite{HT1986}. The engulfment loss cone is in grey and comets engulfed are shown as crosses. The regime on the left of $a_\crit$ is the empty loss cone, on the right is the full loss cone. This figure closely follows Figure 1 in \protect\cite{OConnor2023}.\label{fig:diagram_loss_cone}}
\end{figure}

\cite{HT1986} study the injection rate of Oort cloud comets into the Solar System. To do so, they use the galactic tide equations of motion, together with the loss cone theory framework as proposed by \cite{Lightman1977}. Without perturbations from a planetary companion, this can be used to estimate the rate of Oort cloud comets capable of being excited to any arbitrarily small pericentre distance.

First, we describe Oort cloud comets by a distribution function $f$, defined as the number of comets per volume of phase space:
\begin{equation}\label{eqn:distribution_function}
    \dd N_\mathrm{Oort} = f(\Lambda, L, L_z, \omega, \Omega, l) ~\dd \Lambda ~\dd L ~\dd L_z ~\dd \omega ~\dd \Omega ~\dd l.
\end{equation}
Following previous work \citep{HT1986, WiegertTremaine1999, OConnor2023}, the distribution function is integrated over angles, assuming a spherical distribution of comets. This is motivated by observations in long-term Solar System Oort cloud simulations that the Oort cloud is spherically symmetric \citep[e.g.][]{Duncan1987,Vokrouhlicky2019}.

\begin{align}\label{eqn:dN_f}
    \dd N_\mathrm{Oort} &= f(\Lambda, L) \dd L~ \dd \Lambda~ \int_{-L}^{L} \dd L_z~ \int_0^{2\pi} \dd \omega~ \int_0^{2\pi} \dd \Omega~ \int_0^{2\pi} \dd l \nonumber\\
                        &= (2\pi)^3  f(\Lambda, L) \cdot 2 L ~\dd L~ \dd \Lambda
\end{align}
where $\int \dd N_\mathrm{Oort} = N_\mathrm{Oort}$ with $N_\mathrm{Oort}$ as the total number of comets in the Oort cloud. With this description of the Oort cloud, we can proceed to find the pollution rate.

In the WD pollution case, we are interested in rate of comets that can be excited to $q = q_\crit = 1 R_\odot$, the fiducial Roche limit of a WD that we adopt. At this distance, we assume that a comet will be tidally disrupted by the WD.

For comets with high eccentricity, $e\lesssim 1$, the angular momentum can be related to the pericentre distance $q$ as:
\begin{equation}
    L \approx (2 G M_* q)^{1/2}.
\end{equation}
Through this, we define a critical angular momentum once a comet has achieved a certain critical pericentre distance $q_\crit$:
\begin{equation}
    L_\crit \equiv (2 G M_* q_\crit)^{1/2}.
\end{equation}
The loss cone is defined as the phase space region where $L \leq L_\crit$. 
This is the tidally-disrupted loss cone. Once a comet enters this loss cone, we assume it is removed from the Oort cloud (because it is tidally disrupted).

Comets can be injected into the loss cone in two regimes: the filled and empty loss cones, depending on their semi-major axis $a$ (or equivalently, $\Lambda$).
Intuitively the dependency on $a$ is because these two regimes depend on how strong the galactic tide can affect a comet. 
In one case the galactic tide induces small changes in angular momentum $\Delta L$ over multiple orbits, slowly migrating a comet inward in $q$ over many orbits. This is the empty loss cone case. 
In the other case the galactic tide is sufficiently strong to induce a large $\Delta L$ and the comet is capable of reaching the loss cone within one orbit.
This is the filled loss cone case. Figure \ref{fig:diagram_loss_cone} \citep[following Figure 1 in][]{OConnor2023} is a diagram illustrating these two regimes.

First, we consider the empty loss cone case, which happens when $\Delta L \leq L_\crit$, where $\Delta L \sim |\dd L / \dd t| \times P$ is the angular momentum change induced by the galactic tide per orbit. \cite{HT1986} show that the comet injection rate in the empty loss cone regime, $\Gamma_e$ (dimension of number of comets per unit $\Lambda$ per unit time) is:
\begin{equation}
    \Gamma_e (\Lambda) ~\dd \Lambda = \frac{160 \pi^3 G \rho_0}{3} \frac{L_\crit \Lambda^4}{(G M_*)^2} f(\Lambda, L_\crit) ~\dd \Lambda
\end{equation}
where $L_\crit \ll \Lambda$ (highly eccentric orbit).
$\Gamma_e$ is found by using Equation \ref{eqn:Ldot_GT} and calculate the rate at which comets are pushed into the loss cone boundary at $L_\crit$.

Next, we consider the filled loss cone case when $\Delta L \geq L_\crit$. The injection rate for the filled loss cone, $\Gamma_f$ (same dimension as $\Gamma_e$) is:
\begin{equation}
    \Gamma_f (\Lambda) ~\dd \Lambda = 4 \pi^2 (G M_*)^2\frac{L_\crit^2}{\Lambda^3} f(\Lambda, L_\crit) ~\dd \Lambda
\end{equation}
where $\Gamma_f$ is found by dividing the number of comets inside the loss cone by the comets' orbital period (because these comets are lost within one orbit). Note that the subscript $f$ here denotes `full', not the distribution function $f$.

The loss cone is empty at small $a$ and full at large $a$. The transition between the two cases can be found by equating the two rates, $\Gamma_e = \Gamma_f$, yielding:
\begin{align}\label{eqn:a_crit}
    a_\crit &= \left(\frac{3}{20\sqrt{2}\pi} \frac{M_*}{ \rho_0} q_\crit^{1/2}\right)^{2/7} \nonumber\\
    &\approx 10~500 \mathrm{~AU} \cdot \left(\frac{M_*}{0.6 M_\odot}\right)^{2/7} \left(\frac{\rho_0}{0.1 M_\odot~\mathrm{pc^{-3}}}\right)^{-2/7} \left(\frac{q_\crit}{1 R_\odot}\right)^{1/7}.
\end{align}
When $a < a_\crit$, a comets is in the empty loss cone regime.
When $a \geq a_\crit$, it is in the full loss cone regime.
In Delaunay variables, these conditions are equivalent to $\Lambda < \Lambda_\crit$ and $\Lambda \geq \Lambda_\crit$ with
\begin{equation}
    \Lambda_\crit \equiv (G M_* a_\crit)^{1/2}.
\end{equation}

The total injection rate (number of comets entering a certain $q_\crit$ per unit time) can be found by adding up the two injection rates, integrated over $\Lambda$:
\begin{equation}\label{eqn:Gamma_total}
    \Gamma_\mathrm{total} = \int_{\Lambda_1}^{\Lambda_\crit} \Gamma_e(\Lambda) ~\dd \Lambda ~~+~ \int_{\Lambda_\crit}^{\Lambda_2} \Gamma_f(\Lambda) ~\dd \Lambda
\end{equation}
where $\Lambda_1 = (G M_* a_1)^{1/2}$ and $\Lambda_2 = (G M_* a_2)^{1/2}$, with $a_1$ the inner semi-major axis edge of the Oort cloud and $a_2$ the outer edge. 

Finally, since $\Gamma_e \propto a^2 f(a)$, the injection rate per unit $\Lambda$ increases from $a_1$ to $a_\crit$.
On the other hand, $\Gamma_f \propto a^{-3/2} f(a)$, the injection rate per unit $\Lambda$ decreases from $a_\crit$ to $a_2$. At $a_\crit$, $\Gamma_e = \Gamma_f$.
Since the injection rate increases until $a_\crit$ and then decreases, the majority of the total injection rate is contributed from the region around $a_\crit$.

\subsection{Oort cloud Distribution Function}\label{sec:distribution_function}

Now, we find the explicit form of the distribution function $f$. First, for a dynamically relaxed distribution, the distribution is `thermal' and the distribution in $e^2$ is uniform \citep{Jeans1919, Ambartsumian1937}. This `thermal` distribution of Oort cloud comets is seen after long term simulations of Solar System Oort cloud \citep[c.f. Figure 7 in][]{Vokrouhlicky2019}. In this case, $f = f(\Lambda)$ is independent of $L$. We can further simplify by integrating over $L = \sqrt{G M_* a (1-e^2)}$ which is now uniform between $[0, (G M_* a)^{1/2}] = [0, \Lambda]$:
\begin{align}
    \dd N_\mathrm{Oort} = (2\pi)^3  f(\Lambda) ~\dd \Lambda \int_0^{\Lambda} 2 L ~\dd L~ = (2\pi)^3 f(\Lambda) \Lambda^2 ~\dd \Lambda.
\end{align}

By definition, for a spherically distributed shell with width $[a, a + \dd a]$, we also have:
\begin{equation}\label{eqn:dNda}
    \dd N_\mathrm{Oort} \equiv n(a) ~\dd V = n(a) \cdot  4 \pi a^2 \dd a
\end{equation}
where, $n(a)$ is the number density profile of comets. 

Before finding $f(\Lambda)$ through these equations, we first choose, $n(a)$.
Long term simulations of the Solar System Oort cloud find that our own Oort cloud generally follow a powerlaw density profile \citep[e.g.][]{Duncan1987, Higuchi2015, Vokrouhlicky2019}:
\begin{equation}
    n(a) \propto a^{-\gamma}
\end{equation}
where $a \in [a_1, a_2]$, with $a_1$ and $a_2$ as the inner and outer semi-major axis edge of an Oort cloud.
Following \cite{OConnor2023}, we consider $2 \leq \gamma \leq 4$ in this article, with $\gamma = 3.5$ as a fiducial value for a Solar System-like Oort cloud.

With a description of $n(a)$, we can find the distribution function $f(\Lambda)$ by equating the two relations for $\dd N$ \citep{OConnor2023}:
\begin{equation}
    f = f(\Lambda) = \begin{cases}
                        C N_\mathrm{Oort} (\Lambda / \Lambda_1)^{3-2\gamma} &,\Lambda_1 \leq \Lambda \leq\Lambda_2 \\
                        0 &,\text{otherwise}
                    \end{cases}
\end{equation}
where $\Lambda_1$ and $\Lambda_2$ are the circular angular momenta associated with $a_1$ and $a_2$. $C$ is the normalisation constant:
\begin{equation}
    C = \frac{8 \pi}{(2 \pi)^3} \cdot \begin{cases}
                        (\gamma - 3)\left(4 \pi \Lambda_1^3 (1 - (\Lambda_1 / \Lambda_2)^{2\gamma - 6}\right)^{-1} &,\gamma\neq 3\\
                        \left(8\pi^3 \Lambda_1^3 \ln (\Lambda_2/\Lambda_1)\right)^{-1} &,\gamma= 3
                    \end{cases}.
\end{equation}

Equipped with the distribution function $f(\Lambda)$, it is possible to find the total comet injection rate $\Gamma_\mathrm{total}$ (dimension of number of comets per unit time) by substituting $f(\Lambda)$ and integrating Equation \ref{eqn:Gamma_total}.
With $\Gamma_\mathrm{total}$, we can also predict the pollution rate (i.e. in $\mathrm{g~ s^{-1}}$) into a WD:
\begin{equation}
    \dot{M}_Z = \Gamma_\mathrm{total} \cdot \frac{M_\mathrm{Oort}}{N_\mathrm{Oort}}.
\end{equation}

\section{Analytic Theory: Companion Dynamics}\label{sec:analytic_theory_companion}

In this section, we refer to a companion in a binary system with the WD as planetary if it has mass $M \leq 10^{-2} M_\odot = 10 M_\Jup$, or stellar if it has mass $M \geq 10^{-1} M_\odot$.

In section \ref{sec:precession}, we analyse the effects of companion-induced precession on a comet with semi-major axis $a$ and pericentre $q$ from a companion with mass $M_p$ on a circular orbit at semi-major axis $a_p$. Companion induced precession is compared against galactic tide induced precession. 
In section \ref{sec:precession_efficiency}, we study the efficiency of precession at preventing comets' migration due to galactic tide through simulations. 

\subsection{Precession}\label{sec:precession}

Galactic tidal effects can be suppressed by angular momentum change induced by a companion, which is also accompanied by an apsidal precession.
We compare the apsidal precession rates, $\dot{\omega}$, induced by galactic tide and companion to study when each effect is dominant.
In our case, we consider the regime where $q \geq a_p$ --- no orbit crossings occur and the companion must be interior to the comet.

With secular forcing by a companion, a comet experiences apsidal precession (at the quadrupole order) at a rate \citep{FaragoLaskar2010}\footnote{Gau{\ss} noticed that this is equivalent to the apsidal precession rate of a test particle induced by the quadrupole moment of a homogeneous ring with mass $M_* M_p / (M_* + M_p)$ and radius $a_p$ \citep[see references in][]{Touma2009}.}:
\begin{align}\label{eqn:omegadot_p}
    \langle \dot{\omega}_p \rangle &= \frac{M_p M_*}{(M_* + M_p)^2} \cdot \left(\frac{a_p}{a}\right)^{7/2} \cdot \frac{3 n_p}{8 (1-e^2)^{2}} \cdot \left(5\cos^2 \Delta I - 1\right)
\end{align}
where $n_p$ is the companion's mean motion and $\Delta I$ is the mutual inclination between the comet and companion.
This is a secular interaction, integrated over orbital motions of both companion and comet.
There are higher order terms to $\dot{\omega}_p$ \citep[the next non-zero term occurs at the hexadecapole order, see][]{PalacianYanguas2006, VinsonChiang2018}, which become important as $q\to a_p$.
However, for our analysis in this section, the quadrupolar term is sufficient to give an estimate.
When we require numerical results for $\dot{\omega}_p$ (Figure \ref{fig:timescales}), $\dot{\omega}_p$ is measured numerically and is not limited by this approximation. 

Galactic tide likewise induces apsidal precession (Equation \ref{eqn:omegadot_GT}). Companion-induced precession begins to dominate that of galactic tide when these rates are comparable:
\begin{equation}\label{eqn:dominance_condition}
    \zeta \equiv \abs{\frac{\langle \dot{\omega}_{\mathrm{GT}}\rangle}{\langle\dot{\omega}_p\rangle}} \sim 1.
\end{equation}
Precession in argument of pericentre ($\Delta \omega$) accompanies angular momentum change ($\Delta L$).
That is, the companion also induces a change in angular momentum, suppressing the angular momentum change from galactic tide and inhibiting further migration in pericentre $q$.

Expanding $\zeta$ to first order in $q/a$ (the high eccentricity limit, $e \sim 1$) and taking the order of magnitude terms, we find:
\begin{align}\label{eqn:analytic_Phi}
    \zeta &\sim \frac{32 \pi \sqrt{2}}{3} \cdot \alpha \cdot \beta \simeq 50 \cdot \alpha \cdot \beta
\end{align}
where we defined two dimensionless quantities $\alpha$ and $\beta$. 

$\alpha$ is the ratio of densities between the galaxy and the binary system:
\begin{align}\label{eqn:alpha}
    \alpha &\equiv \frac{\rho_0}{M_\mathrm{reduced} / a_p^3} \nonumber\\ 
            &\approx 10^{-9} \cdot \left(\frac{\rho_0}{0.1 \mathrm{M_\odot~pc^{-3}}}\right)\left(\frac{M_\mathrm{reduced}}{10^{-2} M_\odot}\right)^{-1} \left(\frac{a_p}{100\mathrm{~AU}}\right)^3
\end{align}
where $M_\mathrm{reduced} = M_* M_p / (M_* + M_p)$ is the reduced mass.
As $\alpha$ becomes smaller, $\rho_0 \ll M_\mathrm{reduced} / a_p^3$, we expect the effects of the binary to be stronger than that of galactic tide. 

$\beta$ compares the orbit of a comet to the companion's orbit:
\begin{align}\label{eqn:beta}
    \beta &\equiv \left(\frac{q}{a_p}\right)^{3/2} \left(\frac{a}{a_p}\right)^{7/2} = \left(1-e\right)^{3/2} \cdot \left(\frac{a}{a_p}\right)^5 \nonumber\\
            &\approx 10^{7} \cdot \left( \frac{q}{a_p} \right)^{3/2} \left(\frac{a}{10^4 \mathrm{~AU}}\right)^{7/2} \left(\frac{a_p}{100 \mathrm{~AU}}\right)^{-7/2}.
\end{align}
Higher $\beta$ means that comet experiences less torque from the companion.
$\beta$ is mostly dominated by $a/a_p$. When a comet has a large orbit comparing to the companion ($a \gg a_p$), it spends most of its orbital time far from the companion and thus, receiving less torque. $\beta$ is large in this case. There is also a dependence on $q$ (or equivalently, on the eccentricity $e$). As a comet migrates inward due to galactic tide, the orbit becomes more eccentricity, $e$ increases, $q$ decreases and thus $\beta$ also decreases.

\begin{table}
    \caption{Order-of-magnitude values of the dimensionless quantity $\alpha$, which describes the ratio of densities between the galaxy and the binary system (Equation \ref{eqn:alpha}).}
    \label{tab:alpha}
    \begin{tabular}{lcccc}
        \hline
        $\alpha$ & $M_p$ [$M_\odot$] & $M_*$ [$M_\odot$] & $M_\mathrm{reduced}$ [$M_\odot$] & $a_p$ [AU] \\
        \hline
        $10^{-14}$ & 0.6 & 0.6 & 0.3 & 10 \\
        $10^{-13}$ & $10^{-1}$ & 0.6 & $10^{-1}$ & 10 \\
        $10^{-12}$ & $10^{-2}$ & 0.6 & $10^{-2}$ & 10 \\
        \hline
        $10^{-10}$ & 0.6 & 0.6 & 0.3 & 200 \\
        $10^{-9}$ & $10^{-1}$ & 0.6 & $10^{-1}$ & 200 \\
        $10^{-8}$ & $10^{-2}$ & 0.6 & $10^{-2}$ & 200 \\        
    \end{tabular}
\end{table}

\begin{table}
    \caption{Order-of-magnitude $\beta$ values for comets with pericentre distance $q = 5 a_p$. The dimensionless quantity $\beta$ describes the comet's orbit relative to the companion orbit (Equation \ref{eqn:beta}).}
    \label{tab:beta}
    \begin{tabular}{lccc}
        \hline
        $\beta$ & $a_p$ [AU] & $q$ [AU] & $a [AU]$ \\
        \hline
        $10^{11}$ & 10  & 50 & $10^4$ \\
        $10^{8}$  & 100 & 500 & $10^4$ \\
        $10^{7}$  & 200 & 1000 & $10^4$ \\
    \end{tabular}
\end{table}

\begin{figure}
    \centering
    \includegraphics[width=\the\columnwidth]{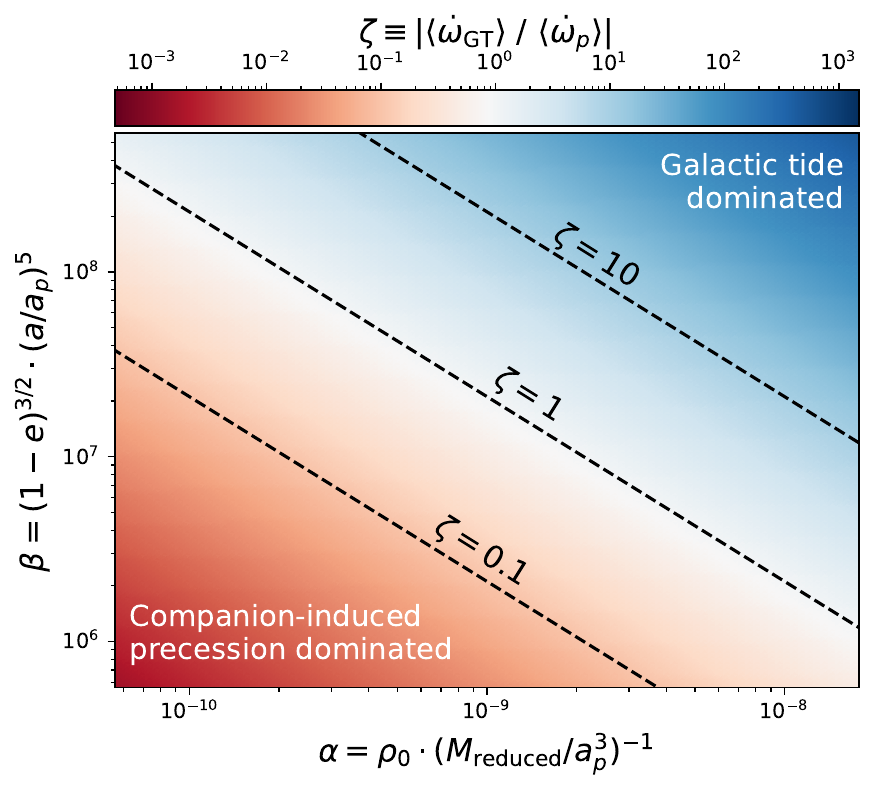}
    \caption{Values of $\zeta$ over a grid of $\alpha$ and $\beta$ (see Table \ref{tab:alpha} and \ref{tab:beta} for typical values). Dashed lines are where $\zeta = 0.1, 1, 10$. In the top right corner, $\zeta \gg 1$ and galactic tide is stronger than companion-induced precession, and vice-versa in the bottom left.
    }    \label{fig:precession_efficiency_on_Phi}
\end{figure}

Typical values of $\alpha$ and $\beta$ are shown in Tables \ref{tab:alpha} and \ref{tab:beta}.
In table \ref{tab:alpha}, typical scenarios are evaluated: WD -- WD binary, WD -- M dwarf, and WD -- large giant planet.
Table \ref{tab:beta} shows cases where the comet pericentre is at $q=5 a_p$.
This is chosen because here, scattering is generally weaker comparing to both precession and galactic tide (see Figure \ref{fig:timescales}).
A comet semi-major axis of $10^4$ AU is chosen because this is typical for incoming comets.
In both tables, we give sample values for a close-in companion at $a_p = 10$ AU and a wider companion at $a_p = 100 - 200$ AU.

Figure \ref{fig:precession_efficiency_on_Phi} shows $\zeta$ on a grid of $\alpha$ and $\beta$. The dashed lines show the contours where $\zeta = 0.1, 1, 10$. 
Near and below $\zeta = 1$, we expect companion-induced precession to be stronger than galactic tide; that is, precession can overcome galactic tide and suppress inward migration.
We confirm the analytic values of $\zeta$ through simulations measuring the torque a comet experiences due to only galactic tide and only companion-induced precession.

For close companions at $a_p = 10$ AU, we have $\beta > 10^{11}$. 
In this case, to have precession dominates tide by making $\zeta \lesssim 1$, we require $\alpha < 10^{-13}$, which can be supplied by a stellar-mass companion with mass $M_p \geq 0.1 M_\odot$ as shown in Table \ref{tab:alpha}. 
However, a planetary-mass companion does not provide sufficient torque to overcome galactic tide. 
For more distant companions at $a_p = 100 - 200$ AU, the same conclusions can be reached through their values of $\zeta$.
Therefore, a stellar-mass companion ($M_p \geq 0.1 M_\odot$) is required to produce a precession barrier, which reduces the efficiency of galactic tide delivering comets from an exo-Oort cloud.

\subsubsection{Limitations}
We note that there are four limitations when using $\zeta$:
\begin{enumerate}
    \item $\dot{\omega}_p$ as shown in Equation \ref{eqn:omegadot_p} is only valid for $q \geq a_p$; that is, a comet must be strictly exterior to companion.
    \item We only use the quadrupole term for $\dot{\omega}_p$. As $q$ approaches $a_p$, contributions from higher order terms become important. Thus, $\zeta$ should not be used when $q \approx a_p$. 
    \item We ignore angular dependencies in $I, \omega, \Omega$. However, as seen in Figure \ref{fig:precession_efficiency} where we averaged results over angles, $\zeta$ without angular dependencies is still able to give an order of magnitude estimate of the strength between precession and galactic tide.
    \item We expand $\zeta$ in the limit $q \ll a$. In other words, $\zeta$ is not a good approximation for very circular orbits ($q \approx a$), or for very close-in comets encountering very widely-separated companions ($a$ is small, but $a_p$ is large, so $q$ is comparable to $a$). For the first case, when $q \approx a$, the comet is on a circular orbit and mostly affected by galactic tide and not by the companion. For the second case, we must limit our analyses to companions with $a_p \ll a$. The closest Oort cloud comets in our model have $a \approx 3000$ AU, so at most, we should only consider $a_p \leq 300$ AU for $\zeta$ analysis.
\end{enumerate}

\subsection{Precession Barrier Efficiency}\label{sec:precession_efficiency}
\begin{figure}
    \centering
    \includegraphics[width=\columnwidth]{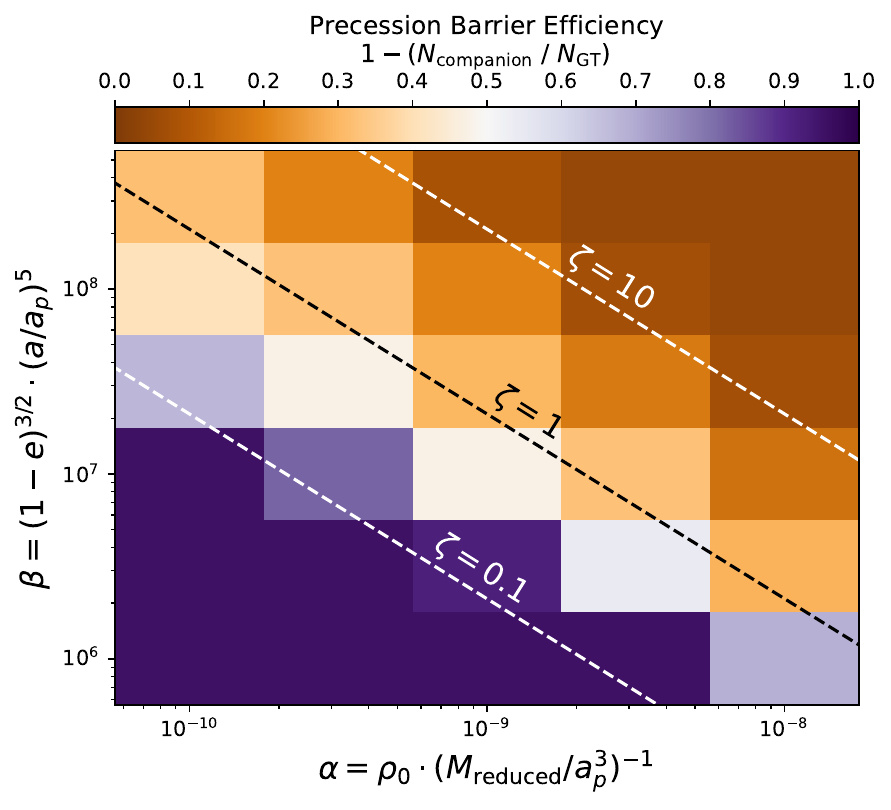}
    \caption{Efficiency of the precession barrier induced by a companion, measured from numerical simulations.
    Simulation values of $M_p, M_*, a, a_p, q$ are chosen from grid values of ($\alpha, \beta$) so that scattering does not matter.
    In other words, this efficiency is purely from effects of the torque induced by a planetary companion reducing the effectiveness of galactic tide.
    The colour shows the efficiency of the planet's torque at inhibiting galactic tide, where $N_\mathrm{companion}$ is the number of comets that can enter a certain $q$ in the existence of a companion and galactic tide and $N_\mathrm{GT}$ is the same but there is only galactic tide (no companion). 
    Dashed lines show contours for $\zeta=0.1, 1, 10$.    
    Above the $\zeta = 10$ dashed line, the precession barrier efficiency is 0\% and comets fully experience galactic tide and can be excited to high eccentricity, migrating inward in pericentre $q$.
    Below the $\zeta = 10$ dashed line, we begin to see the precession barrier suppressing galactic tide and the efficiency decreases.
    At the very bottom left point, galactic tide is completely suppressed, and efficiency is $\sim 100$\%.
    \label{fig:precession_efficiency}}
\end{figure}

Figure \ref{fig:precession_efficiency} presents the efficiency of the precession barrier at overcoming galactic tidal effects over a grid of $\alpha$ and $\beta$.
The efficiency in this figure is defined as:
\begin{equation}\label{eqn:precession_efficiency}
    \mathrm{Efficiency} = 1 - (N_\mathrm{companion} / N_\mathrm{GT})
\end{equation}
where $N_\mathrm{GT}$ is the number of comets that can enter a certain pericentre $q$ with only galactic tide.
$N_\mathrm{companion}$ is the same number, but there are galactic tide and a companion.
As defined, the efficiency measures the effectiveness of a companion in suppressing tidal effects.
A companion can do this either by inducing a precession barrier or by inducing a scattering barrier.
In this subsection, we only consider the efficiency of the precession barrier.

We found $N_\mathrm{GT}$ and $N_\mathrm{companion}$ numerically.
For each set of $(\alpha, \beta)$, two simulations are run: with and without a companion.
In both simulations, galactic tide is included.
This is evaluated on a grid of $\alpha$ and $\beta$, which depends on $M_p, M_*, a, a_p, e$.
These values are chosen so that the effect of scattering is always weaker than that of galactic tide and companion-induced precession.
Other Keplerian orbital elements for a comet $(I, \omega, \Omega)$ are drawn randomly assuming comets are isotropically distributed.

We compare the efficiency over a grid of ($\alpha, \beta$) to the analytic values of $\zeta$ in Figure \ref{fig:precession_efficiency}.
Above the $\zeta = 10$ contour, companion-induced precession does not suppress any comets experiencing galactic tide.
Here, 100\% of comets can come into $q$ due to galactic tide.
As $\zeta$ decreases, the precession barrier becomes stronger and eventually at the bottom left corner, all comets are suppressed from galactic tide, preventing inward migration in $q$. 
At $\zeta \sim 1$, when galactic tide is on the order of companion-induced precession, about half of the comets are suppressed from entering.
Below $\zeta = 0.1$, the barrier efficiency is 100\% and all comets are suppressed from galactic tide.

These efficiency behaviours found from simulations match well with what we expected from $\zeta$ over many orders of magnitude: when $\zeta \lesssim 1$ companion-induced precession suppresses galactic tide.
We find that $\zeta$ is a useful indicator of where companion-induced precession is important relative to galactic tide. 
Specifically, the contour $\zeta \approx 10$ is a good indicator separating regimes of where galactic tide dominates and where precession dominates.
When $\zeta$ decreases, precession begins to dominate galactic tide, and becomes increasingly more effective as $\zeta \ll 1$ where the efficiency tends to 100\%.

\subsection{Scattering Timescale}

Comets not only experience a torque in $\Delta L$ causing precession, but also experience a kick in energy causing a change in semi-major axis.
There are two regimes where comets experience semi-major axis kicks from a massive companion: strong and diffusive scattering.
In the strong scattering regime, $q \gtrsim a_p$, comets are ejected from the system within one pericentre passage.
Here, the scattering timescale is approximately $T_\mathrm{scattering} \approx P_\comet / 2$; comets are kicked during their incoming pericentre passages.

In the diffusive regime, $q > a_p$, comets experience small random semi-major axis kicks during multiple pericentre passages.
These kicks vary in strength due to the phase difference between the comet and the companion during the encounters.
\cite{HaddenTremaine2023} investigate comet-companion interactions through an analytic mapping approach.
Using the results they found, we can write the characteristic timescale on which small diffusive kicks in semi-major axis will lead to an order unity change in binding energy (or equivalently, order unity change in semi-major axis)
\footnote{Equation 25 in \cite{HaddenTremaine2023} is re-written as the timescale on which the comet experiences a strong scattering event due to diffusive kicks in semi-major axis.}
:
\begin{align}
    T_\mathrm{scattering}
    \approx ~1.5 \times 10^4 \mathrm{yr}~\cdot &\left(\frac{M_p}{10^{-2} M_\odot}\right)^{-2}
    \left(\frac{M_*}{0.6 M_\odot}\right)^{3/2}\cdot\nonumber \\ &\left(\frac{a}{10^4 \mathrm{~AU}}\right)^{3/2}\exp\left(7.4 \cdot \frac{q}{a_p}\right).
\end{align}

There are three main assumptions to using the results found by \cite{HaddenTremaine2023} diffusive timescale for Oort cloud comets. 
First, the incoming comet is assumed to be coplanar to the companion-WD orbital plane. 
This is not true for Oort cloud comets, which are isotropically distributed. 
Second, $M_p \ll M_*$, the companion's mass is much smaller than the central star's mass. 
Hence, we cannot use this timescale to estimate the scattering timescale in the stellar-mass companion case.
Third, they assume the comet is on a parabolic orbit ($e=1$).
While this is not exactly true in our case, Oort cloud comets have highly eccentric orbits when they interact with companions.
Hence, we further assume that we can use the $e = 1$ diffusive timescale as written here, for comets with $e \lesssim 1$.
Finally, since we assume $e\lesssim 1$, this timescale is only applicable in cases where $q \ll a$ and $a_p \ll a$ (the companion is much closer-in than the comet).

Next, we compare analytic expectations of $T_\mathrm{scattering}$ to numerical simulations.
We simulate two cases: a stellar-mass companion with $M_p = 0.6 M_\odot$ and planetary-mass companion with $M_p = 10^{-2} M_\odot$.
In both cases, the companions are on a circular orbit at $a_p = 200$ AU around a central WD with mass $M_* = 0.6 M_\odot$ and the companion's initial phase is randomised.
At each pericentre $q$, where $q\in[1, 6]a_p$, 200 comets are initialised with $a = 15~000$ AU (typical semi-major axis of incoming comets to interact with companions at $a_p = 200$~AU).
The comets' initial position is set at apocentre.
Comets' angles are initialised either coplanar and isotropically. Simulations are stopped when comets experience a kick 
$\Delta a / a_0 > 0.3$ relative to the initial semi-major axis $a_0$, or until the simulation time reaches $10^{11}$~years and we call this time the numerical scattering timescale.
The numerical condition, $\Delta a / a_0 > 0.3$, is somewhat arbitrary but we found that this is a good indicator for when comets experience order unity changes in binding energy (strong scattering).

\begin{figure}
    \centering
    \includegraphics[width=\the\columnwidth]{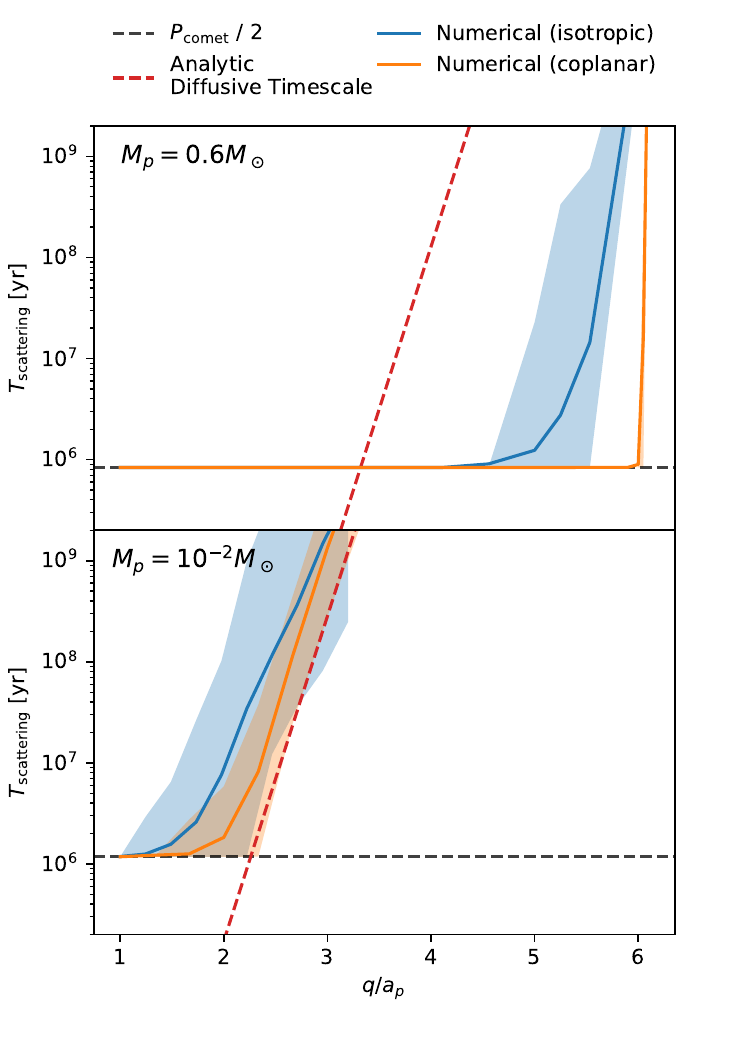}
    \caption{Numerical scattering timescales for coplanar (blue) and isotropic (orange) comets interacting with a stellar-mass (top) and planetary (bottom) companion. The analytic diffusive timescales are shown in both panels, but can only be used in the bottom panel where $M_p \ll M_*$. Blue and orange shaded areas are the range of measured numerical scattering timescales. In both cases, the high diffusive timescale flattens to the strong scattering regime (where comets are ejected within one pericentre passage with timescale $P_\comet / 2$) as $q$ decreases.
    }    \label{fig:scattering_timescales}
\end{figure}

Figure \ref{fig:scattering_timescales} compares numerical scattering timescales with analytic expectations.
In both panels, numerical timescales for coplanar and isotropic comets are shown. 
The solid coloured lines are the mean scattering timescales and the shaded areas are the timescale ranges from 200 comets.
The top panel shows the timescales for a stellar-mass companion and the bottom panel for a planetary-mass companion. 
First, there are clearly two scattering regimes.
Comets sufficiently far away experience long scattering timescale according to the diffusive timescale.
As the pericentre decreases, the timescale eventually converges to the strong scattering regime where comets are ejected within one pericentre passage.
Second, in the bottom panel, we find the timescale for coplanar comets interacting with a planet matches well within an order-of-magnitude with the analytic expectations.
The slopes between numerical and analytic match well in the diffusive regimes.
The analytic $T_\mathrm{scattering}$ is consistently off by a factor of a few.
We attribute this to the arbitrary $\Delta a/a_0 > 0.3$ numerical scattering condition.
Third, isotropic comets have a higher scattering timescale.
In addition, as the pericentre increases, coplanar and isotropic timescales converge.
This behaviour is expected since isotropic comets experience weaker kicks than coplanar comets.
But at high pericentre distances, their kick strengths are both small.
Fourth, in the top panel, we find the analytic expectation no longer gives a reliable scattering timescale.
This is because the condition $M_p \ll M_*$ is strongly violated.
Here, we observe that coplanar comets quickly becoming strongly scattered at $q \approx 6 a_p$, while isotropic comets transition slower.
Note that in an investigation of a test particle on an initially circular orbit around a binary, \cite{Holman1999} found a stability limit around $4 a_p$ for equal mass binaries.
Their result is different from ours because their test particles are on circular orbits while ours are highly eccentric.
Hence, their test particles can experience effects from mean motion resonances, as discussed in detail in \cite{Holman1999}.
That being said, their result and ours are reminiscent of each other.

Once a comet reaches a pericentre within the strong scattering regime, they cannot survive multiple encounters.
Thus, when there is a scattering barrier strong enough to create a strong scattering regime, comets in the empty loss cone are prevented from slowly migrating inward. 
However, this does not mean there are no comets entering smaller $q$.
Comets in the full loss cone regime can still migrate within one orbital period to be engulfed by the WD because they only interact with the companion once.
This is the motivation for the modified loss cone theory in section \ref{sec:modified_loss_cone}.

\subsection{Modified Loss Cone Theory} \label{sec:modified_loss_cone}

\begin{figure}
    \centering
    \includegraphics[width=\the\columnwidth]{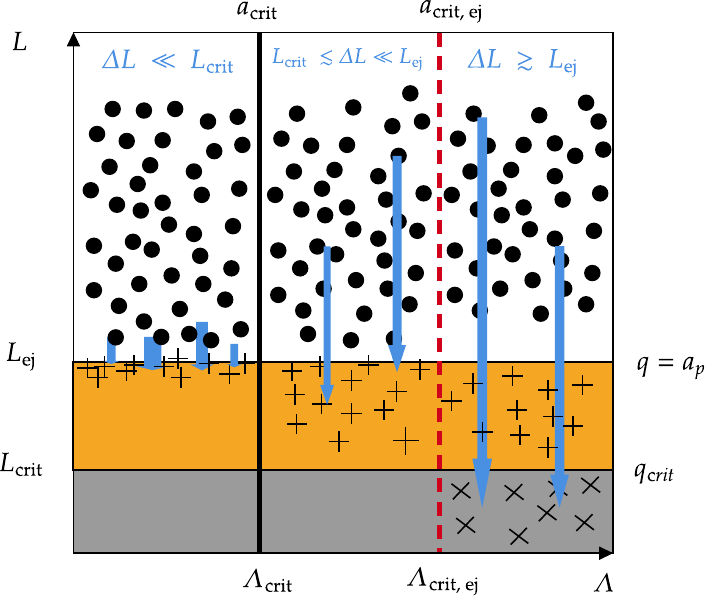}
    \caption{Diagram illustrating the modified loss cone theory by \protect\cite{OConnor2023}. The engulfment loss cone is in grey, the ejection loss cone is in orange. Comets ejected are shown as plus signs and comets capable of engulfment are shown as crosses. This figure is created following closely Figure 12 in \protect\cite{OConnor2023}.\label{fig:diagram_modified_loss_cone}}
\end{figure}

\cite{OConnor2023} estimate the reduction of pollution rate in the presence of a planetary companion by modifying the loss cone theory.
We summarise their theory below and illustrate the modified loss cone theory in Figure \ref{fig:diagram_modified_loss_cone}. 
We also discuss the limitations of this modified loss cone framework.
A similar effect is proposed by \cite{Teboul2024} for dense star clusters with a central black hole called ``loss cone shielding''. 

An `ejection loss cone' is defined as the region where comets are shielded from further migrating inward due to repeated encounters with the planet. 
During multiple strong encounters, the comet experiences changes in semi-major axis which can eventually eject the comet.
The location of the ejection loss cone is defined at:
\begin{equation}
    L_\mathrm{ej} = \sqrt{2 G M_\mathrm{total} a_p}
\end{equation}
nesting on top of the tidal-disruption loss cone at
\begin{equation}
    L_\mathrm{crit} = \sqrt{2 G M_\mathrm{total} q_\crit}.
\end{equation}
where $M_\mathrm{total} = M_* + M_p$ is the total mass of the WD star and the planetary companion.
In the case where the planet is far away from the tidal radius, $a_p \gg q_\crit$, we have $L_\mathrm{ej} \gg L_\mathrm{crit}$.
The two loss cones overlap, but the ejection loss cone covers a much larger region of phase space than the tidal loss cone.

As a comet experiences a change in angular momentum $\Delta L$ due to galactic tide, it drifts inward encountering these loss cones.
A comet with relatively small $a$ experiences a small $\Delta L \ll L_\mathrm{ej}$, and drifts slowly to the edge of the ejection loss cone; the ejection loss cone is empty. 
Vice versa, a comet with larger $a$ experiences a much greater $\Delta L \gtrsim L_\mathrm{ej}$, and jumps through the ejection loss cone in one orbital period; the ejection loss cone is filled.
In addition, if a comet experiences sufficiently large change in angular momentum, $\Delta L \gtrsim L_\mathrm{crit}$, it can jump through both loss cones and can be tidally disrupted and pollute a WD.
These regimes are illustrated in Figure \ref{fig:diagram_modified_loss_cone} \citep[following Figure 12 in][]{OConnor2023}.
The comet avoids ejection since it reaches the engulfment loss cone in one orbital period.
The transition semi-major axis between the two $\Delta L$ regimes is:
\begin{align}
    a_\mathrm{crit, ej} &\sim a_\crit \cdot \left(\frac{a_p}{q_\crit}\right)^{1/7} \nonumber\\
    &\approx 32~000 \mathrm{~AU} \cdot \left(\frac{a_p}{10\mathrm{~AU}}\right)^{1/7} \left(\frac{q_\crit}{1 R_\odot}\right)^{-1/7}
\end{align}
where in the last expression we used values of $a_\crit = 10~500\mathrm{~AU}$ (Equation \ref{eqn:a_crit}), $q_\crit = 1 R_\odot$ and $a_p = 10$ AU.
This transition semi-major axis corresponds to a critical circular momentum
\begin{equation}
    \Lambda_\mathrm{crit, ej} = \sqrt{G M_\mathrm{total} a_\mathrm{crit, ej}}.
\end{equation}

Under this framework, only comets with 
$a > a_\mathrm{crit, ej}$ can experience a sufficiently large $\Delta L \gtrsim L_\mathrm{ej}$ to pollute a WD. In addition, comets that can pollute WDs are in the filled loss cone regime because $a_\mathrm{crit, ej} > a_\mathrm{crit}$. Therefore, the pollution rate is estimated as the filled loss cone rate integrated over semi-major axes ranging between $a_\mathrm{crit, ej} \leq a \leq a_2$, or equivalently over $\Lambda$:
\begin{equation}\label{eqn:planet_theory_rate}
    \Gamma_\mathrm{total, planet} = \int_{\Lambda_{\mathrm{crit, ej}}}^{\Lambda_2} \Gamma_\mathrm{f}(\Lambda) ~\dd \Lambda.
\end{equation}
\cite{OConnor2023} further simplify this expression to find:
\begin{equation}\label{eqn:planet_theory_rate_2}
    \frac{\Gamma_\mathrm{total, planet}}{\Gamma_\mathrm{total, GT}} \sim \left(\frac{q_\crit}{a_p}\right)^{(2\gamma - 1)/14}.
\end{equation}
Taking a fiducial $\gamma = 3.5$ (for a Solar System Oort cloud), the factor on the right hand side is approximately 0.05 for $q_\crit = 1 R_\odot$ and $a_p = 10$ AU.
With this, we analytically expect the existence of a planetary companion to reduce WD pollution rate by about 1.5 orders of magnitude for $\gamma = 3.5$.

\subsubsection{Limitations}\label{sec:modified_loss_cone_limitations}

First, comets with $\Delta L \ll L_\mathrm{ej}$ drift slowly to the edge of the ejection loss cone and then are assumed to be fully ejected when they enter into the ejection loss cone.
However, comet-planet interactions can be weak and not sufficiently strong to eject comets.
For example, we expect a comet to experience a much weaker kick in semi-major axis by a much smaller mass planet.
A planet the size of the Earth and a planet with $10 M_\Jup$ will create loss cone barriers with very different efficiency.
Another example is if a comet is very inclined relative to the planet's orbital plane, the kick will also be much weaker.
Therefore, not every comet with $a < a_\mathrm{crit, ej}$ will be ejected. 
In the small planet mass limit, we expect $\Gamma_\mathrm{total, planet} = \Gamma_\mathrm{total, GT}$ since the planetary ejection barrier is not effective at all.
In the large planet mass limit, the planet creates a 100\% effective scattering barrier, \textit{reducing} pollution rate as described by Equation \ref{eqn:planet_theory_rate_2}.
\cite{OConnor2023} recognised that planet mass should affect ejection efficiency, yet Equation \ref{eqn:planet_theory_rate_2} does not have a mass dependence.
To identify which planet could create a sufficiently strong ejection barrier, they propose a quantity
\footnote{$\lambda$ is called $\Lambda$ in \cite{OConnor2023}. We already used $\Lambda$ in this work for the circular angular momentum.}
:
\begin{equation}\label{fig:lambda_oconnor}
    \lambda = \left(\frac{M_p}{M_\Jup}\right) \left(\frac{M_\odot}{M_*}\right) \left(\frac{a}{10^4 \mathrm{AU}}\right) \left(\frac{10 \mathrm{AU}}{a_p}\right).
\end{equation}
If $\lambda \ll 1$, the ejection barrier is weak and the comet receives negligible kicks and can safely migrate inward through multiple orbits.
If $\lambda > 1$, then the ejection barrier becomes important.
Using $\lambda$, a $10 M_\Jup$ planet at $a_p=10$ AU should be able to create a sufficiently strong ejection barrier.
We will test the ejection barrier strength of a planet with this configuration later.

Second, this theory cannot account for additional dynamics that can be induced by a planetary mass companion.
For example, a comet can be captured into smaller orbits and experiences more complicated dynamics which can also facilitate delivery into WDs, such as von Zeipel-Kozai-Lidov or inverse Kozai \citep{VinsonChiang2018, FaragoLaskar2010}.

Third, as discussed earlier, another important additional dynamics is that a comet experiences small random kicks in semi-major axis, $\Delta a$, through multiple diffusive encounters.
This causes the semi-major axis to change over time which also changes how a comet experiences galactic tide $\Delta L$ over time.
Since $\Delta L$ due to galactic tide is strongly dependent on a comet's semi-major axis, these random walks in $a$ induced by a planet can potentially cause a comet to experience vastly different Galactic tidal effects over time.
Therefore, assuming a WD pollution rate through integrating the total full loss cone rate $\Gamma_\mathrm{f}$ ranging between a fixed range of semi-major axes for all comets over all time (Equation \ref{eqn:planet_theory_rate}) might not be a suitable estimate. 

Consider an example where a comet begins with an initial semi-major axis $a < a_\mathrm{crit, ej}$.
Through galactic tide, the comet migrates inward in pericentre distance $q$.
At some point later in time when the comet achieves a $q \gtrsim a_p$, the comet begins to experience random kicks in semi-major axis every pericentre passages through interactions with the planet. 
Due to multiple small $\Delta a$ kicks, the comet is migrated to $a \gtrsim a_\mathrm{crit, ej}$.
In this example, we initially count this comet to be in the empty ejection loss cone and thus, cannot pollute comets.
But random interactions with a planet bring the comet to a larger $a$, where galactic tide induces a stronger $\Delta L$ allowing the comet to bypass the ejection loss cone and pollute the WD.
In this example, having a larger planet might actually \textit{increase} pollution rate because larger planets can induce stronger random walks in $a$ at a larger range pericentre range.

In the Solar System, this mechanism for Oort cloud comets to bypass the Jupiter-Saturn ejection barrier is shown in \cite{Kaib2009}.
Here, weak perturbations by Uranus and Neptune are attributed to induce small kicks in $a$, bringing comets into a stronger $\Delta L$ regime capable of bypassing the ejection barrier.
As shown in Figure \ref{fig:scattering_timescales}, one single planet can likewise induce small perturbations on comets.
Thus, the comets in our case can similarly bypass the ejection barrier through these small kicks in $a$.

Finally, there is a small chance of a comet becoming unbound through a strong kick, but can still pollute a WD on its last inbound passage.

\subsection{Timescales Comparison}\label{sec:timescales}

\begin{figure}
    \centering
    \includegraphics[width=\the\columnwidth]{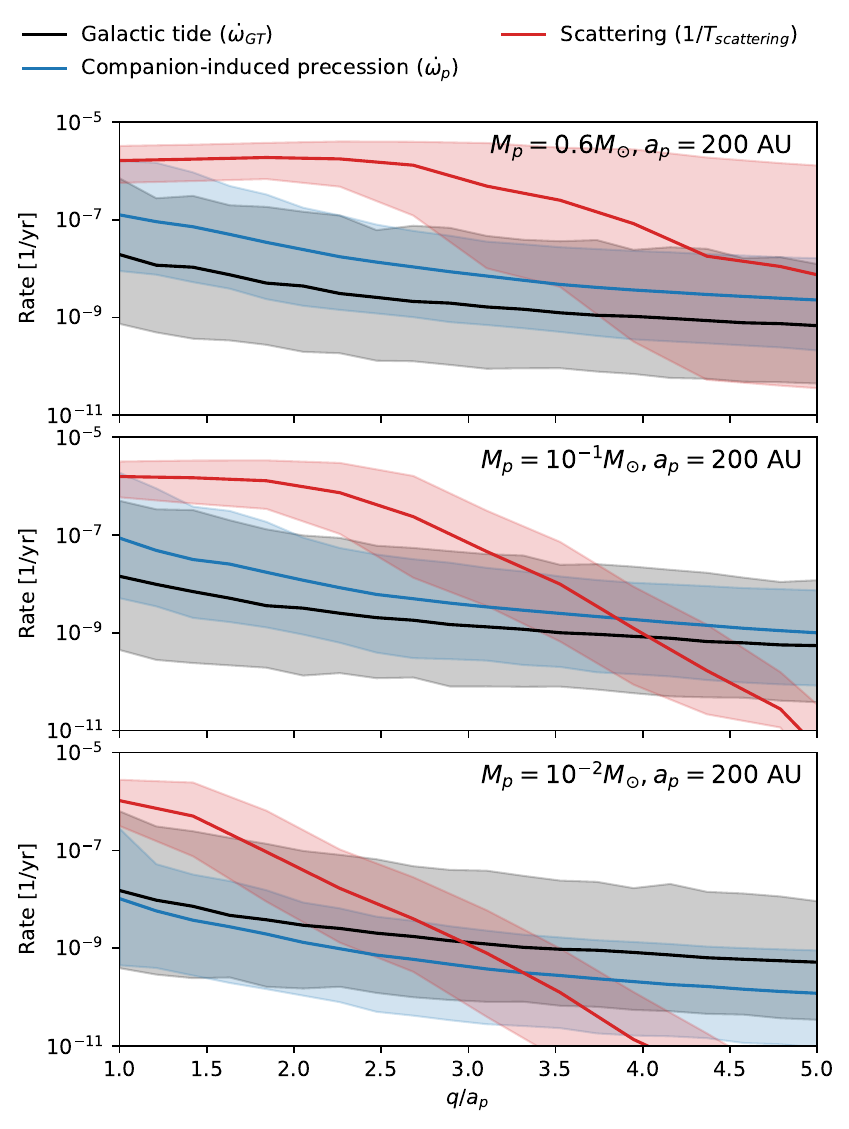}
    \caption{Timescale comparison of galactic tide, companion-induced precession and scattering. From the bottom panel up, these show cases of increasing importance of companions. Timescales are measured from simulations of a Solar System-like Oort cloud ($n(a) \propto a^{-3.5}$, $a \in [3~000, 10^5]$ AU). The shaded area shows the $3\sigma$ spread of timescales for this particular distribution of comets.
    }
    \label{fig:timescales}
\end{figure}

To analyse the importance of the companion's scattering and precession versus galactic tide, we investigate the timescales on which these effects are important.

These timescales are measured from simulations of $10^5$ comets from a Solar System-like Oort cloud: $n(a) \propto a^{-3.5}$ from 3000 to $10^5$ AU.
Here, we directly simulate all comets with an additional galactic tidal force at every timestep --- no secular integration of galactic tide as done in Section \ref{sec:tide_secular_integration}.
This is to ensure that we accurately capture all companion-comet interactions at all $q$.
In addition, we measure independently the contribution of galactic tide and companion at a given pericentre $q$.

First, we simulate $10^5$ comets around a centre-of-mass point mass (the total mass of a WD and its companion) with galactic tide.
This simulation is run until comets' pericentres reach a certain $q$.
The orbital elements of incoming comets with pericentre $q$ are then saved.
Next, we measure the galactic tidal torque over one orbit at this $q$, giving us $\dot{\omega}_\mathrm{GT} \approx \Delta \omega / P_\comet$.
Second, we run another simulation on these saved comets, but now there is a companion and galactic tide is turned off.
In this simulation, we measure the companion-induced precession change on the comet over one orbit, giving us $\dot{\omega}_p \approx \Delta \omega / P_\comet$.
Third, these comets are again run in a third simulation with a companion and no galactic tide to measure the scattering timescale.
The difference between the second and third simulations is in the third simulation comets are simulated longer.
In the third simulation, comets are integrated until they experience a scattering event, $\Delta a / a > 0.3$, or until the simulation time reaches $10^{12}$ years.
Recording the times when comets experience a scattering event gives us $T_\mathrm{scattering}$.
These simulations are run for every $q$ between $q = a_p$ to $q = 5 a_p$ and for companions with masses $M_p = 10^{-2}, 0.1, 0.6 M_\odot$ (corresponding to a WD, M dwarf, and the upper-limit of a planetary mass companion, $10^{-2} M_\odot = 10 M_\Jup$).

We compare the rates (inverse  timescales) of these effects in Figure \ref{fig:timescales}.
The shaded area is the $3\sigma$ spread, showing the range of most comets' timescales.
These three cases show the increasing importance of companion-induced precession and scattering.
As expected, increasing the companion mass increases the strength of scattering.
With a $10^{-2} M_\odot$ planet companion, scattering begins to dominate at $q \approx 3 a_p$, whereas this is increased to $q \approx 3.5 a_p$ for a $0.1 M_\odot$ stellar companion.
For the WD-WD binary case, scattering seems to be dominated in all $q$ considered here.
Note that the scattering timescale flattens out as $q \to a_p$.
This is similar to the scattering behaviour we analysed earlier: comets are ejected within one orbital period in this regime, $T_\mathrm{scattering} \approx P_\comet / 2$.
Similar to scattering, precession's strength becomes stronger relative to galactic tide as companion mass increases.
Finally, Figure \ref{fig:timescales} shows that in all cases, as $q \simeq a_p$, the dominant effect is companion-induced scattering.

\section{Simulation Method}\label{sec:simulation_method}

\begin{figure*}
    \centering
    \tikzset{every picture/.style={line width=1pt}}
    
    \begin{tikzpicture}[x=0.75pt,y=0.75pt,yscale=-1,xscale=1]

    \coordinate (COM1) at (180, 200);
    \coordinate (comet1) at (180, 200+150);

    \draw[black] (COM1) circle (150);
    \node [star, star points=5, star point height=0.05cm, star point ratio=1.6, black, fill=black, draw] at (COM1) {};
    \draw (COM1) -- (comet1);
    \draw[black, fill=newred] (comet1) circle (5);

    \draw (120,154) node [anchor=north west][inner sep=0.75pt]   [align=left] {Centre-of-mass \\(WD and companion)};
    \draw (190,252) node [anchor=north west][inner sep=0.75pt]   [align=left] {$\displaystyle q\ \gg a_{p}$};
    \draw (180-15,200+150+10) node [anchor=north west][inner sep=0.75pt]   [align=left] { \textcolor{newred}{Comet}};

    \draw (5,5) node [anchor=north west][inner sep=0.75pt] [align=left] {\textbf{(1)}};
    \draw (25,5) node [anchor=north west][inner sep=0.75pt]   [align=left] {\textbf{Galactic tide only}: Integrate comet using galactic\\ tide equations of motion up to 1 Gyr.};

    \coordinate (COM2) at (625, 150);
    \coordinate (companion2) at (625, 150-30);
    \coordinate (comet2) at (625-111+150, 150);

    \node [star, star points=5, star point height=0.05cm, star point ratio=1.6, black, fill=newblue, draw] at (625,150) {};

    \draw[gray,dotted] (COM2) circle (30);
    \draw [black, fill=newyellow] (companion2) circle (5);
    \draw (625-50,150-50) node [anchor=north west][inner sep=0.75pt]   [align=left] {\textcolor{newyellow}{Companion}};
    \draw (COM2) -- (companion2);
    \draw (625-25, 150-20) node [anchor=north west][inner sep=0.75pt]   [align=left] { $\displaystyle \ a_{p}$};
    
    \draw (625-111,150) circle [dashed, x radius=150, y radius=100];
    \draw [black, fill=newred] (comet2) circle (5);
    \draw (625+10, 150-15) node [anchor=north west][inner sep=0.75pt]   [align=left] {$\displaystyle q$};
    \draw (COM2) -- (comet2);
    \draw (625-111-40,150-20) node [anchor=north west][inner sep=0.75pt]   [align=left] {$\displaystyle q\ =4a_{p}$};

    \draw (350, 5) node [anchor=north west][inner sep=0.75pt]   [align=left] {\textbf{(2)}};
    \draw (370, 5) node [anchor=north west][inner sep=0.75pt]   [align=left] {Extract comet and go to step \textbf{(3)}\\if pericentre distance reaches $\displaystyle q\ =4\ a_{p}$.};

    \coordinate (COM3) at (340, 430);
    \coordinate (companion3) at (340, 430-25);
    \coordinate (comet3) at (340+149-180, 430);

    \node [star, star points=5, star point height=0.05cm, star point ratio=1.6, black, fill=newblue, draw] at (COM3) {};

    \draw[gray,dotted] (COM3) circle (25);
    \draw [black, fill=newyellow] (companion3) circle (5);
    \draw (COM3) -- (companion3);
    \draw (340+5, 430-20) node [anchor=north west][inner sep=0.75pt]   [align=left] { $\displaystyle \ a_{p}$};
    
    \draw (340+149,430) circle [dashed, x radius=180, y radius=100];
    \draw (340-15, 430+10) node [anchor=north west][inner sep=0.75pt]   [align=left] {$\displaystyle q$};
    \draw [black, fill=newred] (comet3) circle (5);
    \draw (COM3) -- (comet3);
    \draw (340-149+150+70,430-25) node [anchor=north west][inner sep=0.75pt]   [align=left] { $\displaystyle 1\ R_{\odot } \leq \ q\ \leqslant \ 4\ a_{p}$};

    \draw (5,392) node [anchor=north west][inner sep=0.75pt] [align=left] {\textbf{(3)}};
    \draw (25,392) node [anchor=north west][inner sep=0.75pt]   [align=left] {\textbf{Direct integration with REBOUND:}};
    \draw (25,415) node [anchor=north west][inner sep=0.75pt]   [align=left] {Includes galactic tide, a central star\\and a companion.};
    \draw (25,445) node [anchor=north west][inner sep=0.75pt]   [align=left] {Extract comets when comet is engulfed at \\$\displaystyle q= 1\ R_{\odot }$ (Roche radius)};

    \end{tikzpicture}

    \caption{
    A diagram to illustrate our hybrid integration scheme. The first stage is using secular galactic tide equations of motion in (Equations \ref{eqn:Ldot_GT}, \ref{eqn:omegadot_GT} and \ref{eqn:Omegadot_GT}) to quickly integrate a comet. If a comet can be excited to $q = q_\mathrm{switch} = 4 a_p$, then its orbital elements are extracted to be integrated directly using \texttt{REBOUND}. The direct integration by \texttt{REBOUND} uses the full equations of motion and can include interactions with the WD and its companion. A comet is extracted when its apocentre exceeds the WD's Hill sphere ($Q > 0.8$ pc), when it's ejected due to interactions with a companion ($a < 0$) or when it is engulfed by the WD ($d \leq 1 R_\odot$, where $d$ is the distance between the comet and the WD). Integrations are stopped when the simulation time reaches $t=1$ Gyr.
    }
    \label{fig:diagram_hybrid_integration}
\end{figure*}

The previous sections provide us with some understanding of what dynamical effects we should expect in this kind of system.
We now turn to numerical simulations to study the long term (1 Gyr) dynamics of Oort cloud comets under the influence of galactic tide and a companions. There are two main components in our integration scheme: first is a secular integration of the galactic tide equations of motion. If a comet can be excited to a certain $q = q_\mathrm{switch}$, its orbital elements are extracted and the comet is then integrated directly with \texttt{REBOUND} \citep{ReinLiu2012} when $q \leq q_\mathrm{switch}$. When comets violate their boundary conditions (section \ref{sec:boundary_conditions}) they are removed. Figure \ref{fig:diagram_hybrid_integration} illustrates the components of our integration.

\subsection{Integration Scheme}\label{sec:tide_secular_integration}

The majority of Oort cloud comets at most times do not have sufficiently high eccentricity to interact with the central WD or its close companion.

In this regime, the dynamics of a comet is largely governed by its orbit around the WD+companion centre of mass and tidal effects from the galaxy. In this case, we can use the orbit-averaged equations of motion (Equations \ref{eqn:Ldot_GT}, \ref{eqn:omegadot_GT} and \ref{eqn:Omegadot_GT}) to quickly evolve comets. We use the fourth-order Runge-Kutta 
to evolve these coupled differential equations with an adaptive timestep scheme as implemented in \texttt{scipy} \citep{2020SciPy}. We find that the Runge-Kutta adaptive timestep scheme reproduces well the evolution in $(L, \omega)$ as seen in \citet{HT1986}.

If a comet can reach a certain critical $q_\mathrm{switch}$, it is removed from the secular integration and integrated with \texttt{REBOUND} to allow for full interactions with the WD and its companion.
When switched over to integrating with \texttt{REBOUND}, we develop a fast simulation method where only one particle is simulated to further speed up simulation.
This fast \texttt{REBOUND} integration method is illustrated in Figure~\ref{fig:diagram_fast_integration} and described in details in Appendix \ref{sec:direct_integration}.

One caveat of using these orbit-averaged equations of motion is that they are not appropriate when $\Delta L \gg L_\crit$. This is in the filled loss cone regime. In this case, within one orbital period, a comet experiences a significant change in it's angular momentum and are lost. Thus, orbit-averaging is no longer appropriate. This is also resolved by switching over to \texttt{REBOUND} where we integrate with the full (not orbit-averaged) equations of motion.

When a comet achieves a pericentre distance $q_\mathrm{switch} \approx 4 a_p$, interactions with a planetary mass companion become important.
We show this earlier in Figure \ref{fig:timescales} and its discussion.
At $q_\mathrm{switch}$, a comet is extracted from secular equations of motion integration and simulated in \texttt{REBOUND}.
In \texttt{REBOUND}, we simulate full interactions between the comet and WD-companion system, and we use the full (not orbit-averaged) equations of motion for vertical galactic tide.
The companion is set on a circular orbit at a semi-major axis $a_p$ with $I_p=0$. 

When there is no planet, we still extract comets from the secular integration when the comets reach $q_\mathrm{switch} = 10$ AU. This choice is somewhat arbitrary, but it ensures that $q_\mathrm{switch} \gg q_\crit = 1 R_\odot$. This is to make sure that engulfment into the WD is integrated with the full (not orbit-averaged) equations of motion through \texttt{REBOUND}, so that all galactic tidal dynamics are properly captured. As discussed, the secular equations of motion fails in the regime where $\Delta L$ is sufficiently strong to inject the comet into the loss cone in one orbital period.

\begin{figure*}
    \centering
    \tikzset{every picture/.style={line width=1pt}}
    
    \begin{tikzpicture}[x=0.75pt,y=0.75pt,yscale=-1,xscale=1]

    \coordinate (COM) at (250,120);
    
    \draw[gray,dotted] (COM) circle (10);
    \node [star, star points=5, star point height=0.05cm, star point ratio=1.6, black, fill=gray, draw] at (250-10,120) {};

    \draw[gray,dotted] (COM) circle (30);
    \draw [black, fill=gray] (250+15, 120-26) circle (5);
    \draw (250-16,120-45) node [anchor=north west][inner sep=0.75pt]   [align=left] {Companion};
    
    \draw (250+130,120) circle [dashed, x radius=200, y radius=75];
    \draw [black, fill=newred] (250+130+200, 120) circle (5);
    \draw (250+130+200-50+9,120-10) node [anchor=north west][inner sep=0.75pt]   [align=left] {\textcolor{newred}{Comet}};

    \draw (-80,70) node [anchor=north west][inner sep=0.75pt]   [align=left] {\textbf{Fast integration with {REBOUND}}};
    \draw (-80,90) node [anchor=north west][inner sep=0.75pt]   [align=left] {Only 1 particle (the comet) in simulation};
    \draw (-80,110) node [anchor=north west][inner sep=1.5pt]   [align=left] { Force terms are analytically calculated and\\ added to the comet's acceleration at each timestep};

    \draw (390,110) node [anchor=north west][inner sep=0.75pt]   [align=left] { $\displaystyle F_{\text{comet}} = F_{*} +F_{p} + F_\mathrm{GT}$};

    \end{tikzpicture}
    \caption{A diagram illustrating our fast, properly time-adaptive integration method in \texttt{REBOUND}.
    There is only one particle in the simulation: the comet (red).
    The positions of the WD and its companions (shown in grey) can be calculated analytically because they are in a 2-body system.
    After each simulation timestep, the forces are analytically computed and added to the acceleration of the comet particle.
    \texttt{IAS15} is used as an adaptive timestep and is capable of resolve timesteps as: large timestep when the comet is far away (the timestep is a fraction of the comet's orbital period), and small timestep when the comet is close in (fraction of the companion's orbital period).
    This ensures that all close-encounters are properly handled while still maintaining fast integration speed.
    For further descriptions of the fast \texttt{REBOUND} integration method, see Appendix \ref{sec:direct_integration}.}
    \label{fig:diagram_fast_integration}
\end{figure*}
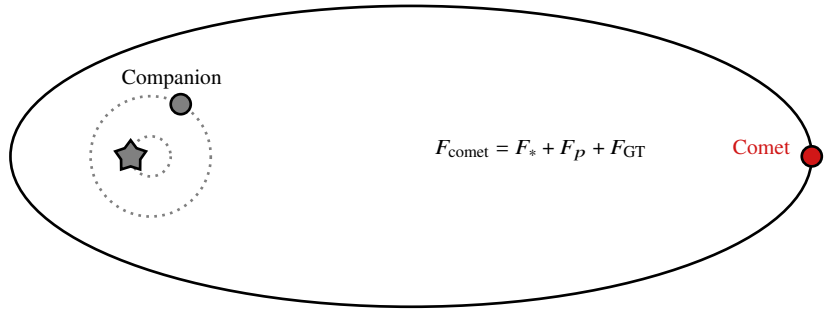

\subsection{Boundary Condition}\label{sec:boundary_conditions}

During the \texttt{REBOUND} part of our simulation, we enforce the following boundary conditions:
\begin{enumerate}
    \item comets with apocentre exceeding the Hill sphere of a $0.6 M_\odot$ star ($Q > 0.8$ pc) and are outbound away from the WD,
    \item comets are ejected ($a < 0$) and are outbound away from the WD,
    \item comets are engulfed ($d \leq 1 R_\odot$, where $d$ is the distance from a comet to the WD)
\end{enumerate}
Only boundary condition (iii) contributes to WD pollution. Boundary conditions (i) and (ii) remove comets because these comets are ejected from the Oort cloud reservoir. 
These three boundary conditions are checked at every simulation timestep. 
In addition, the simulation is stopped when the simulation time reaches $t=1$ Gyr.

The Hill sphere of a WD is scaled down from the Solar System's Hill sphere at $\sim 1$ pc \citep[e.g.][]{Higuchi2015}. For a 0.6 $M_\odot$ WD, the Hill radius is at $0.8$ pc \citep{OConnor2023}. In addition, boundary conditions (i) and (iii)
\begin{align}
    q = a(1-e) &\leq 1 R_\odot \nonumber\\
    Q = a(1+e) &\leq 0.8 \mathrm{~pc}
\end{align}
imply that there is a maximum semi-major axis:
\begin{align}
    a_2 \approx 85~000 \mathrm{AU}.
\end{align}

This is the outer semi-major axis edge of the Oort cloud. 
Comets beyond $a_2$ cannot be excited to high eccentricity since they will be removed for exceeding the apocentre limit.
The lowest pericentre comets at this semi-major axis without exceeding the WD Hill sphere is $q \sim 5~000$ AU.
In addition, this upper semi-major axis limit can also be found by scaling down the Solar System's Oort cloud outer semi-major axis edge at $a_2 = 10^5$ AU, for a $M_* = 0.6 M_\odot$ central star.

\subsection{Initial Condition and Sampling}\label{sec:sampling}

The initial conditions of Oort cloud comets are generated based on a spherical cloud distribution. Comets' argument of pericentre $\omega$ and longitude of the ascending node $\Omega$ are drawn randomly from $\mathcal{U}[0, 2 \pi]$ where $\mathcal{U}$ is the uniform distribution. Comets' inclination are drawn according to $\cos I \sim \mathcal{U}[0, 1]$. The sign of the comet's $I$ is not important since we set the planet at $I_p=0$.

Next, the semi-major axis is drawn such that Oort cloud comets follow a powerlaw density profile $n(a) \propto a^{-\gamma}$:
\begin{equation}
    \dd N(a) \propto a^{-\gamma} a^2 \dd a
\end{equation}
with $a \in [a_1, a_2]$. $a_1 = 3~000$ AU is the inner semi-major axis edge, set based on the Solar System's Oort cloud simulations \citep[e.g.][]{Duncan1987,Vokrouhlicky2019}. $a_2 = 85~000$ AU is the outer semi-major axis edge, as found in the previous subsection.

To efficiently simulate all $\gamma$ without re-running simulations many times, we use a rejection sampling scheme to draw samples for the semi-major axis distribution.
This is described in Appendix \ref{appendix:rejection_sampling}.

After having $a$, we draw the squared eccentricity, $e^2$, from:
\begin{equation}
   e^2 \sim \mathcal{U}[0, 1 - 2 q_\mathrm{initial, min} / a]
\end{equation}
for a distribution uniformly filling the energy phase space \citep{Heisler1990}, appropriate for a dynamically relaxed Oort cloud as seen after long-term simulations of Solar System Oort cloud formation \citep[e.g.][]{Higuchi2015, Vokrouhlicky2019}.
Note that we impose an upper $e$ (or equivalently, a minimum initial pericentre $q_\mathrm{initial, min}$). 
$q_\mathrm{initial, min}$ set sufficiently far that a comet's interaction with a companion or the WD is negligible initially.
This is to ensure that all comets-companion-WD interactions are induced by galactic tide, rather than by random initial condition.
Specifically, in the case where a companion exists, we set
\begin{equation}
    q_\mathrm{initial, min} = 6 a_p.
\end{equation}

In the case where there is only galactic tide and a central WD, it is somewhat more arbitrary:
\begin{equation}
    q_\mathrm{initial, min} = 15 \mathrm{~AU}.
\end{equation}
This is to ensure that $q_\mathrm{initial, min} \gg q_\crit = 1 R_\odot$.
This ensures all comets will be initially integrated using the secular equation of motion and then switched over to be integrated by \texttt{REBOUND} until WD engulfment, ejection, or reaching 1 Gyr.

\section{Numerical White Dwarf Pollution Rate}\label{sec:wd_pollution}

\subsection{Galactic Tide Only}\label{sec:galactic_tide_rate}

\begin{figure}
    \centering
    \includegraphics[width=\the\columnwidth]{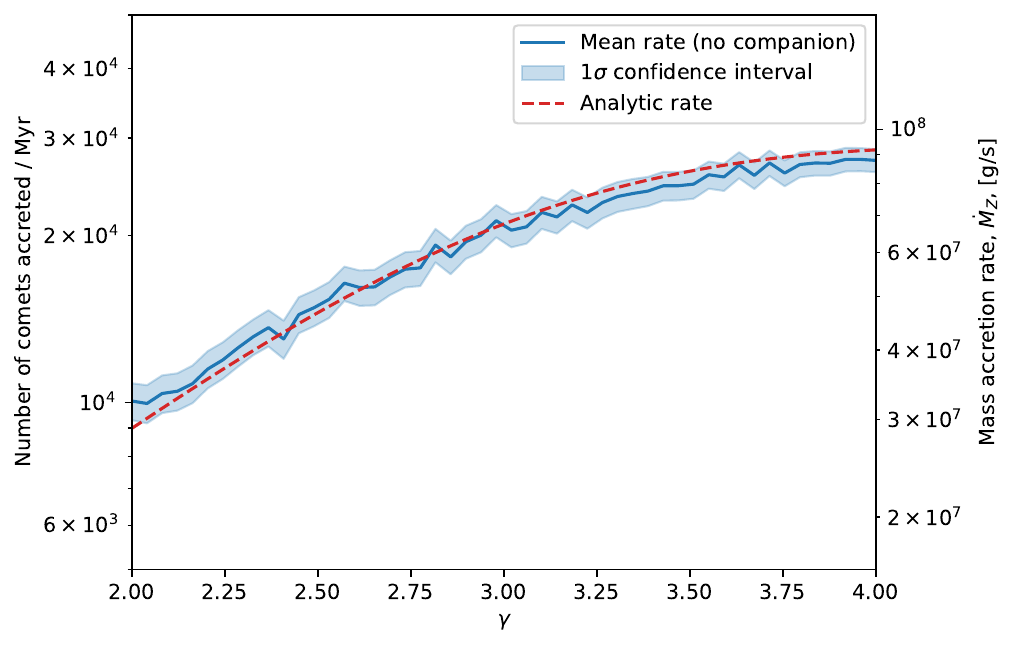}   
    \caption{Accretion rate of comets into $q_\crit = 1 R_\odot$ over various Oort cloud structures $\gamma$, where $n(a) \propto a^{-\gamma}$. The mean simulation rate is shown in blue and the analytic expectation (Equation \ref{eqn:Gamma_total}) is in red. This accretion is due solely to the effects of galactic tide (no companion in this case).  
    The equivalent mass accretion rate on the right is found by assuming an Oort cloud with $N_\mathrm{Oort}=10^{11}$ comets and $M_\mathrm{Oort} = 2 M_\oplus$. The blue shaded area shows the Poisson $1 \sigma$ confidence interval of the solid blue line (mean rate).
    }
    \label{fig:tide_rate_powerlaw}
\end{figure}

We perform numerical simulations to explore the pollution rate of comets into a WD's Roche limit ($1 R_\odot$) over various Oort cloud powerlaw structure ($\gamma$).
Initial conditions and sampling are done according to Section \ref{sec:sampling}, allowing us to sample a variety of powerlaw exponents $\gamma$. 
Here, we simulate with $N_\mathrm{comets}^\mathrm{sim} = 4 \times 10^7$ comets. Because there are no companions in this case, comets are allowed to freely migrate inward in pericentre distance due to galactic tide. 

First, we compare simulation rate of comet accretion with analytic prediction over various Oort cloud structures. 
The rates are shown in Figure \ref{fig:tide_rate_powerlaw}; the solid blue line represents simulation rates and the red-dashed line shows analytic expectations (Equation \ref{eqn:Gamma_total}).
$\gamma$ on the x-axis is the Oort cloud powerlaw exponent value; the number density of Oort cloud objects scale as $n(a) \propto a^{-\gamma}$.
The rate on the left is the number of accreted comets per Myr.
On the right axis is the rate in $\mathrm{g~s^{-1}}$, assuming a fiducial Solar System Oort cloud mass and number of comets with $N_\mathrm{Oort} = 10^{11}$ comets and a total cloud mass of $M_\mathrm{Oort} = 2 M_\oplus$.
To get this rate, we count the total number of comets accreted, $N_\mathrm{accreted}$, after the warm-up phase which is about 400 Myr (subsection \ref{sec:pollution_over_time}).
Dividing $N_\mathrm{accreted}$ by the remaining 600 Myr of simulations yields the number of comets accreted over time as shown on the left hand side of Figure \ref{fig:tide_rate_powerlaw}.

First, we find that simulated rates match well with the analytic expectations from \cite{OConnor2023} based on the framework by \cite{HT1986}.

Second, the average pollution rate of comets into WDs due solely to galactic tide is $\dot{M}_Z \approx 5\times 10^7 - 10^8~\mathrm{g\cdot s^{-1}}$, depending on $\gamma$. Over the course of 1 Gyr, these rates correspond to the delivery $\sim 5 \times 10^{-4} M_\oplus$ of materials. Thus, in the case of galactic tide alone where the only comet removal mechanism is engulfment by the WD, the Oort cloud reservoir is minimally depleted.

Finally, the $1 \sigma$ blue shaded area is found by assuming that comet engulfment is a Poisson process: comet accretion into a WD is a discrete event and comets arrive independently. 
Since we can count the total number of comets accreted, $N_\mathrm{accreted}$, the Poisson process assumption allows us to estimate the standard deviation to be $\sigma = \sqrt{N_\mathrm{accreted}}$. 
Recall that $N_\mathrm{accreted}$ is the total number of accreted comets counted over 600 Myr.
Thus, $\sqrt{N_\mathrm{accreted}}$ is the uncertainty of comets entering $1 R_\odot$ due to our limited number of comets in our simulation ($10^7$ comets) over a $600$ Myr timescale.
In addition, we simulate a sufficiently large number of comets such that $N_\mathrm{accreted}$ is not a small integer and the $1\sigma$ interval can be meaningfully interpreted.
Finally, $\sqrt{N_\mathrm{accreted}}$ is based on the total number of comets in our simulation ($\sim 10^7$ comets).
If our simulation contained $10^{11}$ comets like a full Oort cloud, the blue area would be smaller by 2 orders of magnitude.
In summary, the blue shaded interval is the uncertainty of the total number of accreted comets, $N_\mathrm{accreted}$, from our simulation containing $\sim 10^7$ comets, over a 600 Myr timescale.

\subsection{Efficiency of Companion-induced Precession and Scattering}

\begin{figure}
    \centering
    \includegraphics[width=\the\columnwidth]{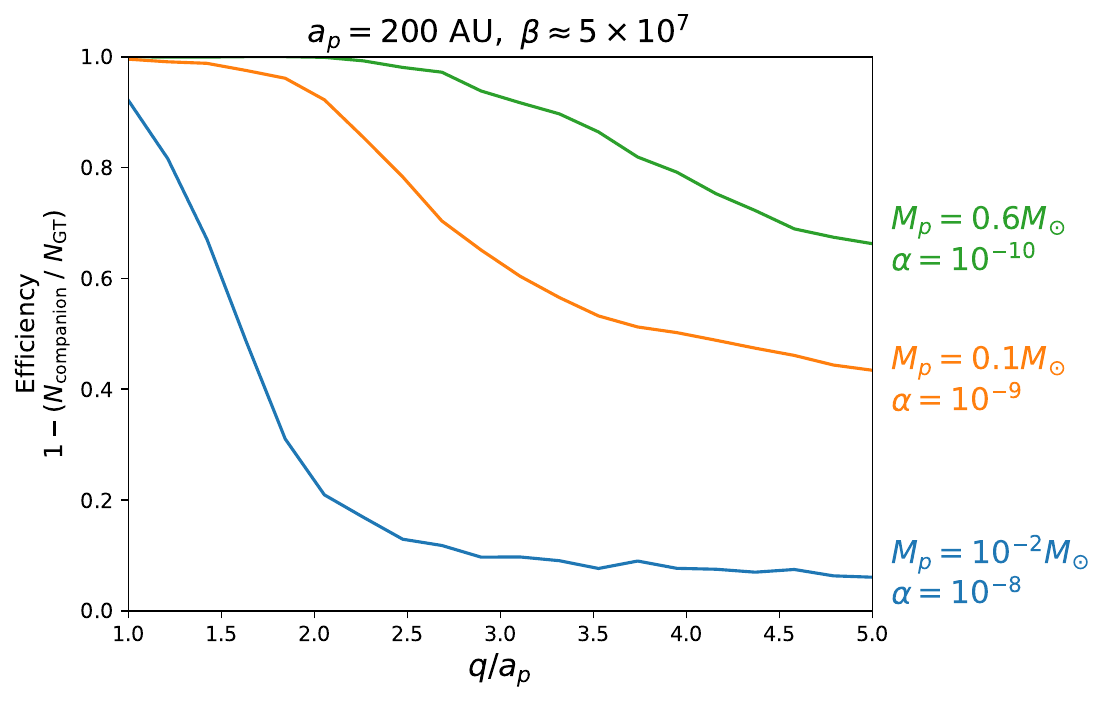}
    \caption{Efficiency of a companion at suppressing galactic tide and preventing comets from entering $q$. These are numerical results from simulations of incoming Solar System-like Oort cloud comets. The three cases here are the same as those in Figure \ref{fig:timescales}. The values of $\alpha$ for each case is shown on the right side with their corresponding colours. At $q=5 a_p$, a typical value for $\beta$ is $\sim 5\times10^{7}$.}
    \label{fig:stellar_efficiency_over_q}
\end{figure}

We now add a companion to our simulation.
We investigate the effects of different companion masses on a Solar System-like Oort cloud.
Figure \ref{fig:stellar_efficiency_over_q} shows the efficiency (Equation \ref{eqn:precession_efficiency}) over different $q$ with varying companion mass: $M_p = 0.6, 10^{-1}, 10^{-2} M_\odot$.
In contrast to Figure \ref{fig:precession_efficiency} previously, we now measure over a range of $q$.
Hence, this efficiency can include the effects of both the precession and scattering barriers produced by a companion.
Especially as $q \to a_p$, both precession and scattering barriers can become very important in increasing the efficiency.
Efficiency is 0\% when galactic tide is the only dominant effect.
Vice-versa, the efficiency is 100\% when precession and scattering barriers induced by a companion can suppress all galactic tidal effects.

First, we use the formulations of $\zeta$ to predict if the precession barrier is stronger than galactic tide at $q=5 a_p$. 
We choose to focus at $q= 5 a_p$ because scattering is not important there for $M_p = 10^{-2}$ and $0.1 M_\odot$ (c.f. timescales in Figure \ref{fig:timescales}).
This can also be done at other $q$ provided those $q$ are within the limitations of our formulations, and scattering is not important.
The three companion cases in Figure \ref{fig:stellar_efficiency_over_q} correspond to $\alpha = 10^{-10}, 10^{-9}, 10^{-8}$. 
From simulations, typical incoming comets into a pericentre $q = 5 a_p = 1000$ AU have semi-major axes $a \sim 15,000$ AU, giving a $\beta \approx 5\times10^{7}$.
For $\alpha = 10^{-8}$, $\zeta = 25 \gg 1$ so we expect galactic tide to be dominant.
For $\alpha = 10^{-9}$, $\zeta \approx 1$ so we expect companion-induced angular momentum change to be important and suppress some comets' galactic tidal torque.
For $\alpha = 10^{-10}$, the precession-barrier is dominant over galactic tide.
Figure \ref{fig:timescales} at $q=5 a_p$ confirms these expectations.

Second, we can predict the efficiency of the precession barrier in reducing comet engulfment by using Figure \ref{fig:precession_efficiency} with values of ($\alpha$, $\beta$).
At $\beta \approx 5\times10^{7}$, Figure \ref{fig:precession_efficiency} predicts about a 0\% efficiency for $\alpha=10^{-8}$.
At $\alpha = 10^{-9}$, it is expected to be around 40\% efficient.
At $\alpha = 10^{-10}$, we expect about a 60\% efficiency.
We confirm these predictions with Figure \ref{fig:stellar_efficiency_over_q} at $q=5 a_p$.

Third, scattering becomes important at $q=3 - 5 a_p$ as seen in the timescale analysis in Figure \ref{fig:timescales}. 
In Figure \ref{fig:stellar_efficiency_over_q}, this corresponds to the fast increase in efficiency at that $q$ range.
In both cases, a stellar-mass companion increases the efficiency by almost 100\% by $q=1 a_p$ due to strong precession and scattering barriers.

Fourth, combining precession and scattering effects, we predict that WDs with a stellar-mass companion are unlikely to be able to be polluted by an Oort cloud exterior to the companion.
The actual rate of WD pollution rate in the presence of a stellar-mass companion is discussed later in Section \ref{sec:WD-WD}.
This is because the WD-star binary case is special due to the centre-of-mass of the system being far from the WD itself.
Thus, to pollute a WD in the presence of a stellar companion, we need to also consider the efficiency of direct collisions between incoming comets and the WD itself.
In contrast, with a planet companion, the centre-of-mass is close to the central WD and all comets migrated to $q \sim 1 R_\odot$ will be engulfed by the WD.

Finally, for a planetary-mass ($M_p \leq 10 M_\Jup$) companion, planet-induced precession does not play a strong role.
At $q = 5 a_p$, the efficiency is still roughly 0\%.
Thus, effects from the planet at the $q$ distance is not important.
There is a quick increase in efficiency beginning at $q \approx 2 a_p$ due to scattering.
However, this is increase is not as strong as the cases with stellar-mass companions to completely prevent further pollution.
We will further analyse the pollution rate in the presence of a planetary-mass companion in the next subsection.

\subsection{Planetary-Mass Companion}

\begin{figure}
    \centering
    \includegraphics[width=\the\columnwidth]{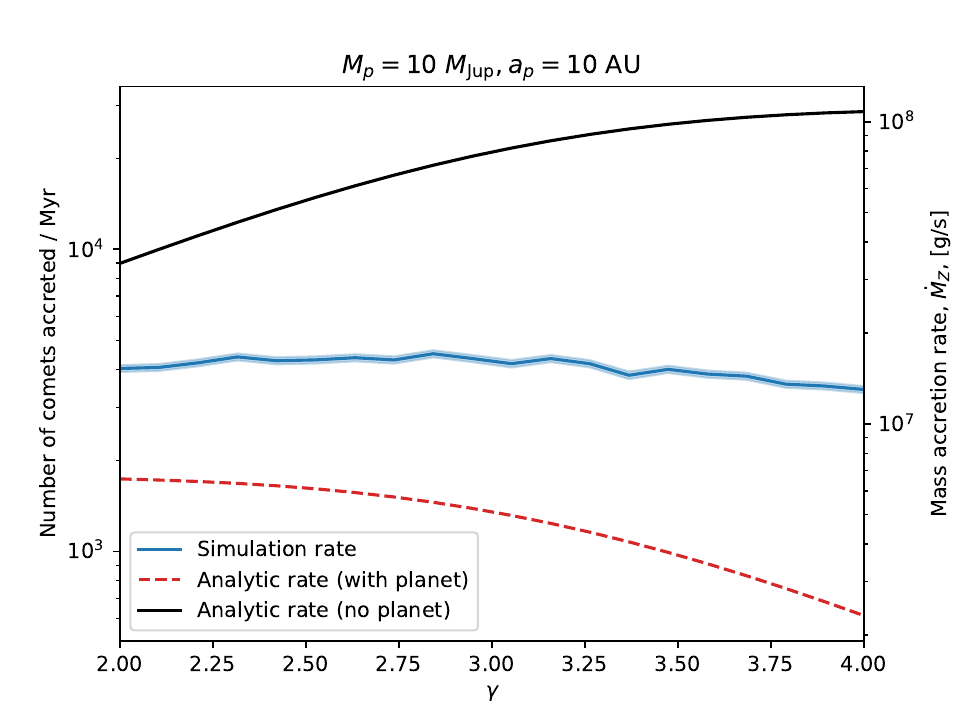}
    \caption{Accretion rate of comets into $q_\crit = 1 R_\odot$ over various Oort cloud structures $\gamma$, where $n(a) \propto a^{-\gamma}$. In this simulation, there is a $10 M_\Jup$ planetary companion at 10 AU. The mean simulation rate is shown in blue and the analytic expectation is in red. The analytic rate in the case where there is no planet (Equation \ref{eqn:Gamma_total}) is shown as a solid black line for comparison. The analytic prediction in the presence of a planet that can create a sufficiently strong ejection barrier (Equation \ref{eqn:planet_theory_rate}) is shown as a red dashed line. The blue shaded area shows the Poisson $1 \sigma$ confidence interval of the solid blue line (mean rate).
    }    \label{fig:tide_rate_powerlaw_planet}
\end{figure}

As discussed in Section \ref{sec:modified_loss_cone}, additional effects due to having a planetary companion are complicated:
having a planet can decrease pollution rate due to the ejection loss cone, but can also potentially increase pollution rate since comets can diffuse to higher semi-major axes to experience stronger galactic tide.
We study the effects of having a planet through numerical simulations and compare with analytic expectations from Section \ref{sec:modified_loss_cone}.
The simulation here uses the same methodology as discussed in Section \ref{sec:simulation_method} with $N_\mathrm{comets}^\mathrm{sim} = 10^8$ comets.

In Figure \ref{fig:tide_rate_powerlaw_planet}, we present the simulated WD pollution rate at various $\gamma$ in the presence of a $M_p = 10 M_\Jup = 10^{-2} M_\odot$ planet at $a_p = 10$ AU. 
These planet mass and semi-major axis values are chosen because this configuration gives a value of $\lambda \approx 10$ (Equation \ref{fig:lambda_oconnor}), which is predicted to create a strong ejection barrier.
In this figure, we find that having a planet decreases the pollution rate into $q_\crit = 1 R_\odot$ by about 1 order of magnitude.
Assuming a Solar System Oort cloud with $N_\mathrm{Oort} = 10^{11}$ comets and a total cloud mass of $M_\mathrm{Oort} = 2 M_\oplus$, the WD pollution rate in this case is $\dot{M}_Z^\mathrm{(planet)} \sim 10^7~\mathrm{g~s^{-1}}$.
We further observe that the simulation rate does not match analytic expectations.
As seen in the plot, simulation rate is about 2-5 times higher than predicted analytic rate.
Furthermore, simulation rate matches better to analytic expectation at low $\gamma$ than at high $\gamma$. 

First, we discuss the behaviour of the analytic expectation.
We notice that the analytic rate decreases as $\gamma$ increases.
This can be understood intuitively through analysing where comets are distributed relative to $a_\mathrm{crit, ej}$.
In the $\gamma=4$ limit, comets are more centrally distributed. Thus, more comets have $a < a_\mathrm{crit, ej}$.
Recall that these are the comets that experience small $\Delta L < L_\mathrm{ej}$ due to galactic tide and migrate slowly in $q$ until they are ejected through encounters with the planet.
Therefore, since most comets are centrally distributed, they have $a < a_\mathrm{crit, ej}$ and we analytically expect most comets to be ejected, reducing the pollution rate.
On the other hand, in the $\gamma=2$ limit, comets are less centrally distributed, some still have $a < a_\mathrm{crit, ej}$ but not as many as in the case of $\gamma=4$.
Therefore, we have more comets with $a > a_\mathrm{crit, ej}$. These comets are capable of experiencing a strong $\Delta L \gtrsim L_\mathrm{crit}$, drifting through the ejection loss cone in one orbit, polluting the WD.
In other words, comet ejection is more effective at $\gamma=4$ than at $\gamma=2$ because there are more comets available to be ejected.

Second, we discuss why simulation rate matches with theory better at $\gamma=2$ than at $\gamma=4$.
With the same intuition where comets are distributed, we further consider that the ejection loss cone is not 100\% effective.
If the barrier is 100\% effective, expect all comets with $a < a_\mathrm{crit, ej}$ to be all ejected.
However, because the ejection loss cone is not 100\% effective, some comets are capable of drifting through the ejection loss cone and eventually pollute the WD.
Since there are more comets at $\gamma=4$ with $a < a_\mathrm{crit, ej}$ than at $\gamma=2$, the assumption of having a 100\% effective ejection loss cone leads us to overestimate the reduction of pollution rate more at $\gamma=4$ than at $\gamma=2$.

\begin{figure}
    \centering
    \includegraphics[width=\the\columnwidth]{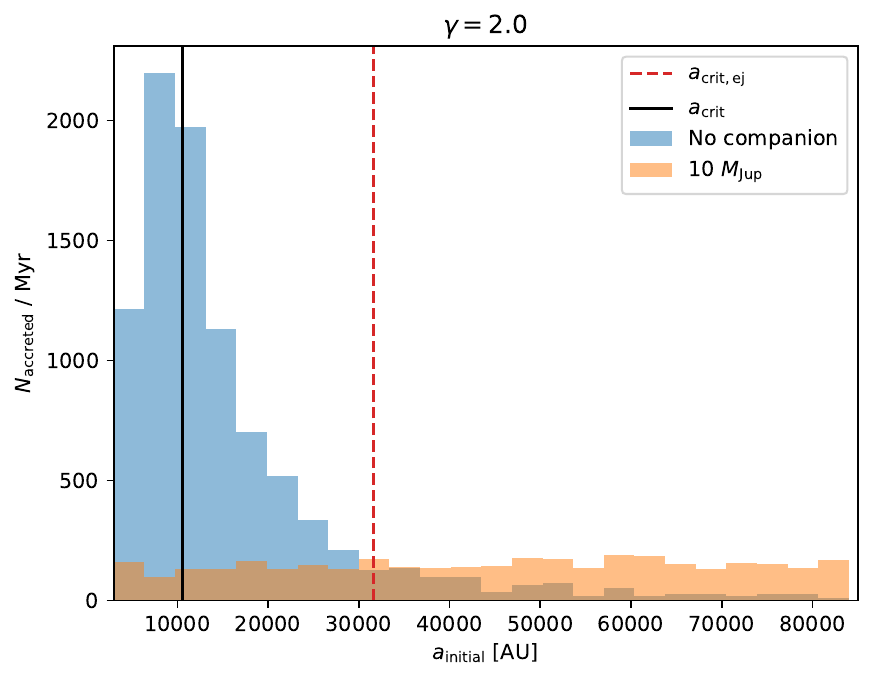}
    \caption{Distribution of accreted comets' initial semi-major axis at $\gamma=2$. Histograms are counted per $a_\mathrm{initial}$ bin. The blue histogram shows distribution when there is galactic tide only. The orange histogram shows the case in the case with galactic tide and a $10 M_\Jup$ planet. Vertical lines show $a_\mathrm{crit, ej}$ (dashed) and $a_\mathrm{crit}$ (solid).
    }    
    \label{fig:N_over_a_planet_2.0}
\end{figure}

Next, we analyse the distribution of accreted comets' initial semi-major axes, $a_\mathrm{initial}$ in Figure \ref{fig:N_over_a_planet_2.0}. 
The initial semi-major axis is shown on the x-axis because comets experience kicks in $a$ over time. 

First, in the regime where $a_\mathrm{initial} < a_\mathrm{crit, ej}$, the planet significantly reduces pollution rate as predicted.
However, we still see comets with $a_\mathrm{initial} < a_\mathrm{crit, ej}$ polluting the WD.
We find that the ejection loss cone barrier is not 100\% effective.
Therefore, the assumption in Equation \ref{eqn:planet_theory_rate} that the pollution rate does not have any contribution with comets from $a < a_\mathrm{crit, ej}$ gives an underestimate of the pollution rate.

Second, in the regime where $a_\mathrm{initial} \leq a_\mathrm{crit, ej}$, the number of comets capable of polluting the WD increases steadily. 
As $a_\mathrm{initial} \to a_\mathrm{crit, ej}$, comets need a smaller kick in $\Delta a$ to deliver them over into the $a \geq a_\mathrm{crit, ej}$ regime where galactic tide can induce a strong enough $\Delta L$ to drift them through the ejection loss cone.
On the other hand, in the $a_\mathrm{initial} \ll a_\mathrm{crit, ej}$, comets will need to experience multiple interactions with the planet that increases their semi-major axis, but not strong enough to eject them.

Third, as $a$ increases beyond $a_\mathrm{crit, ej}$, the pollution rate is higher than in the case of no planet. 
There are several mechanisms to explain this.
First, these comets can be kicked into higher $a$, allowing stronger $\Delta L$ to migrate further in, as described in Section \ref{sec:modified_loss_cone_limitations}.
Second, these comets have lower energy (because of large $a$) and can be kicked into much smaller orbits.
At that point, they can be excited to high eccentricity and pollute the WD with effects like von Zeipel-Kozai-Lidov or inverse Kozai.
Third, they have lower energy, and are also easier to be ejected.
However, in their last inbound passage, they have pericentre distances sufficiently low to pollute the WD.
None of these additional dynamics would be possible without perturbations to the comet.
Without additional perturbations like a planet, a comet at $a=50~000$ AU for example, will stay there and if it cannot reach $q_\crit$ during a galactic tide cycle, will never be able to do so.
Hence, when we assume in Equation \ref{eqn:planet_theory_rate} that the pollution rate strictly follows $\Gamma_\mathrm{f}$ based on a fixed distribution of $a_\mathrm{initial}$, a lot of these additional dynamics are ignored giving an incorrect rate in that regime.
As we have seen, the contribution in region $a > a_\mathrm{crit, ej}$ is higher than expected in Equation \ref{eqn:planet_theory_rate} leading to another source of underestimation of pollution rate.

In summary, we find that the existence of a planetary mass companion significantly reduces the pollution rate for comets with initial semi-major axes $a_\mathrm{initial} \leq a_\mathrm{crit, ej}$ as predicted by \cite{OConnor2023}. However, we find that this reduction is not 100\% effective, that comets experience rich dynamics, and that beyond $a_\mathrm{initial} > a_\mathrm{crit, ej}$ the pollution rate does not simply follow the full loss cone rate $\Gamma_\mathrm{f}$.
That being said, when comparing the overall rates in Figure \ref{fig:tide_rate_powerlaw_planet}, the analytic predictions by \cite{OConnor2023} still yield a good order of magnitude estimate for the pollution rate, although it can be off by a factor of 2-5 times.
We find that the WD pollution rate in the presence of a planetary companion is reduced by one magnitude to $\dot{M}_Z^\mathrm{(planet)} \sim 10^7~\mathrm{g~s^{-1}}$, assuming a Solar System Oort cloud.

\subsection{WD-WD Binary}\label{sec:WD-WD}

\begin{figure}
    \centering
    \includegraphics[width=\the\columnwidth]{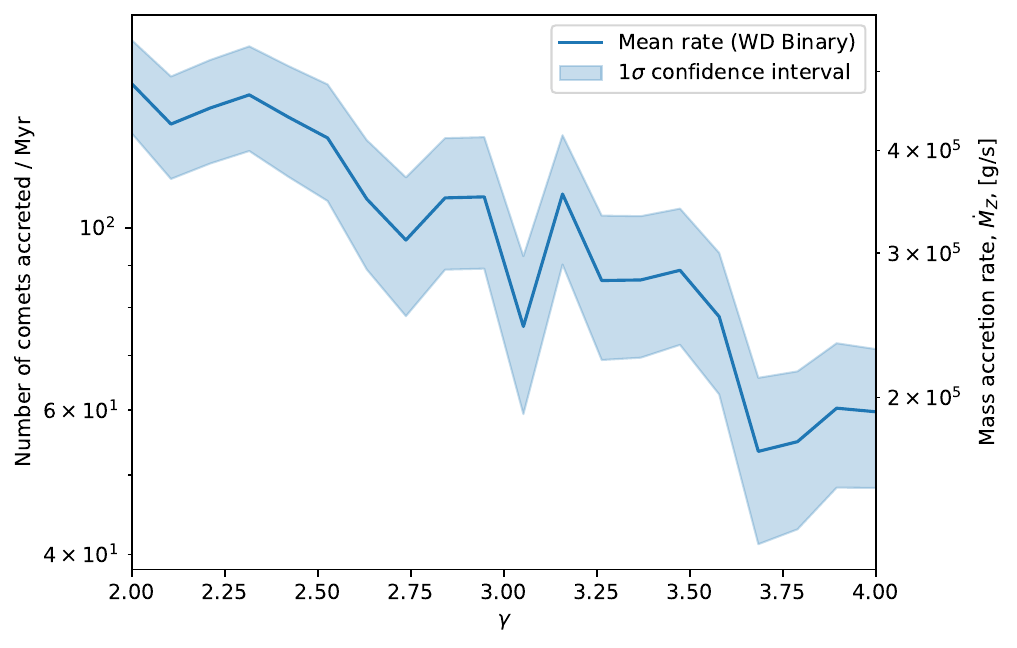}
    \caption{Accretion rate of comets into one WD in a WD-WD binary over various Oort cloud structures $\gamma$. The binary is in circular orbit and separated by $10$ AU. The pollution rate in $\mathrm{g~s^{-1}}$ is shown on the right axis, assuming a Solar System Oort cloud.}
    \label{fig:rate_wd_wd}
\end{figure}

An interesting question is how a WD-WD binary would be polluted by an Oort cloud reservoir.
Unlike the planetary companion case, when the companion's mass is comparable to the WD, the WD is very far from the centre-of-mass.
In this scenario, pollution is governed by direct collisions between incoming comets and the WDs on their orbits.

Assume we have two WDs, separated by $a_p$ on circular orbits.
The Safronov number can be used to estimate how probable it is to expect collisions of incoming comets on highly eccentric orbits:
\begin{equation}
    \Theta = \frac{v_\mathrm{esc}^2}{v_c^2} = 2 \cdot \frac{M_p}{M_* + M_p} \cdot \frac{a_p}{R} \approx 2 \times 10^3
\end{equation}
where $v_\mathrm{esc}$ is the escape velocity from the companion's surface (at distance $R$) and $v_c$ is the circular speed at the companion's semi-major axis.
We set $R = 1 R_\odot$ for the tidal radius where comets are captured by a WD and $a_p = 10$ AU.
Since $\Theta \gg 1$, we predict that collisions into the tidal radius are unlikely \citep{Tremaine2023}.

In Figure \ref{fig:rate_wd_wd}, we show the pollution rate of comets into one WD in a WD-WD binary separated by $a_p = 10$ AU at various $\gamma$.
In the simulation here, we directly check every timestep if a comet's distance is within a WD's tidal radius.
The switching point between secular integration and direct integration with \texttt{REBOUND} is increased to $q=6 a_p$ to accurately capture all dynamics between comets and a stellar mass companions (see Figures \ref{fig:scattering_timescales}, \ref{fig:timescales}).
Here, we find that the pollution rate is significantly reduced by 2.5-3 orders-of-magnitude compared to the galactic tide only case.
Assuming a Solar System Oort cloud, the pollution rate is about $2-4 \times 10^5~\mathrm{g~s^{-1}}$.
Note that this is just below the detection limit at $\sim 5 \times 10^5 ~\mathrm{g~s^{-1}}$.
This rate is low due to a combination of effects.
First, as shown earlier, $\zeta \sim 0.1$ and comets experience a precession torque reducing the effectiveness of galactic tide in exciting incoming comets.
Second, comets in the empty loss cone cannot slowly migrate inwards as they are strongly scattered through multiple encounters with the strong scattering barrier.
Third, comets that can reach sufficiently low $q \approx a_p$ are still unlikely to collide with the WD since the Safronov number is very high.

Next, we consider two extreme cases: a very close WD-WD binary and a more widely separated one. 
If the WD-WD separation is smaller, we expect the pollution rate to increase, peaking at half of the normal galactic tide only rate.
For example, take the limit where the WDs are separated by only a few solar radii, then they are both near the centre-of-mass and will be impacted by all incoming comets.
Since there are two stars, the pollution rate (from the galactic tide only case) will be reduced by half as comets are equally likely to collide with either stars.
If the WD-WD separation is larger, the Safronov number, $\Theta \propto a_p$, would increase and collisions would be even more unlikely.
Thus, we expect a decrease in pollution rate as $a_p$ increases.

Finally, the results here also give an order-of-magnitude estimate of Oort cloud comet pollution rate for other stellar-mass companions, assuming $M_p \sim M_*$.

\subsection{Pollution Over Time}\label{sec:pollution_over_time}

\begin{figure}
    \centering
    \includegraphics[width=\the\columnwidth]{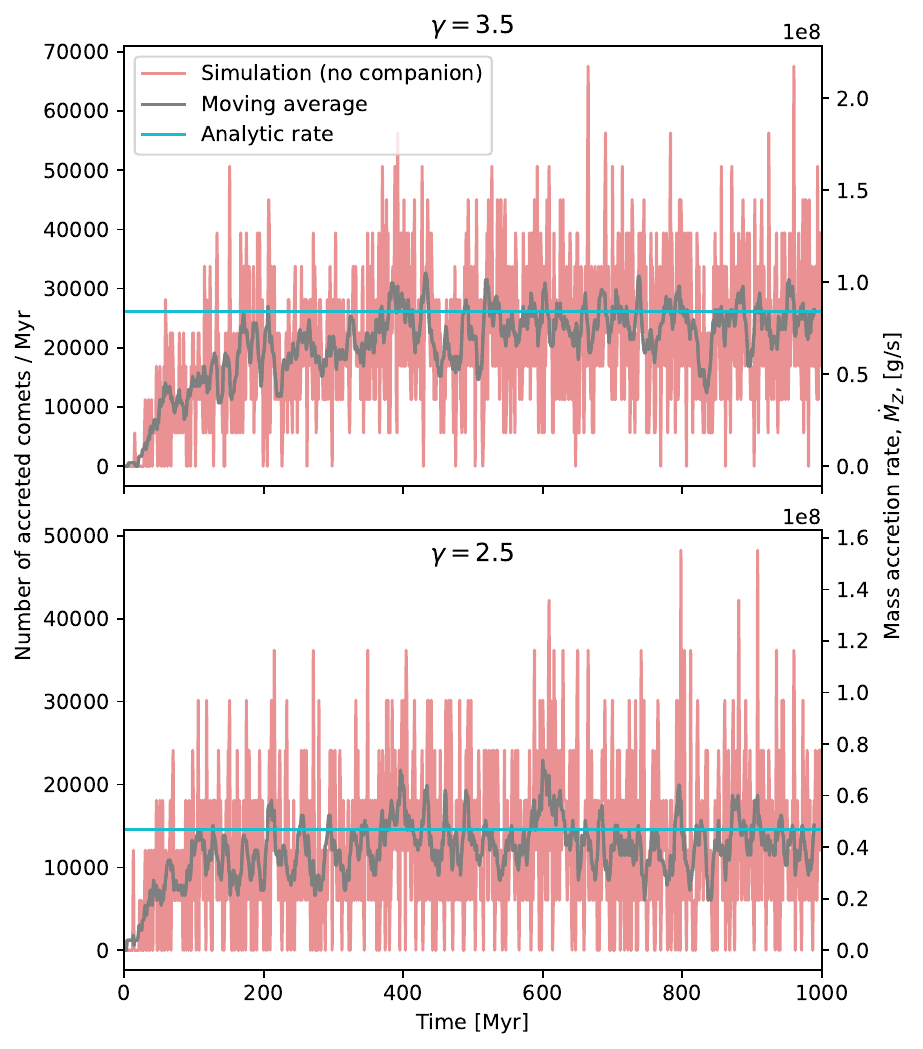}
    \caption{Accretion rate over a 1 Gyr timescale from Oort clouds with $\gamma = 2.5$ and 3.5. There are no companions in this case. Accretion rates from simulations (red) is binned per Myr.  Analytic rates are shown as horizontal blue lines. A moving average with a sliding window of 10 Myr is shown in grey. The accretion rate can be sustained over a 1 Gyr time period. An equivalent mass accretion rate on the right is found by assuming and Oort cloud with $N_\mathrm{Oort} = 10^{11}$ comets and total cloud mass $M_\mathrm{Oort} = 2 M_\oplus$. }
    \label{fig:rate_over_time}
\end{figure}

\begin{figure}
    \centering
    \includegraphics[width=\the\columnwidth]{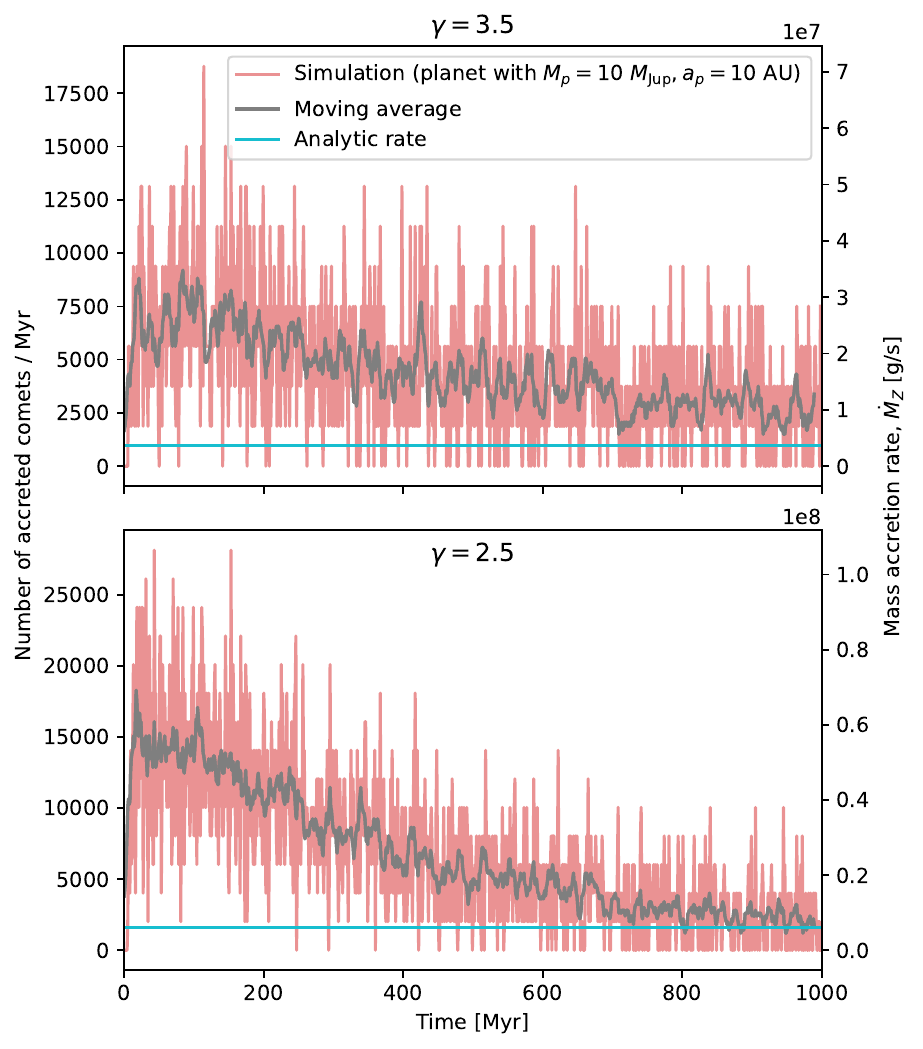}
    \caption{Accretion rate over a 1 Gyr timescale from Oort clouds with $\gamma = 2.5$ and 3.5. There is a planetary-mass companion with $M_p = 10 M_\Jup$ at $a_p = 10$ AU. Accretion rates from simulations (red) is binned per Myr. Analytic expectations are shown as horizontal blue lines. A moving average with a sliding window of 10 Myr is shown in grey. The accretion rate is slightly decreased over the 1 Gyr period, but no more than 0.5 dex.}
    \label{fig:rate_over_time_planet}
\end{figure}

In this subsection, we analyse the pollution rate over time to study if the Oort cloud as a reservoir can consistently maintain WD pollution rate over a Gyr timescale.

In Figure \ref{fig:rate_over_time}, we show the pollution rate over time for $\gamma=$ 2.5 and 3.5 in the case of galactic tide only (no companion).
During the first $\sim$ 400 Myr, we observe a ``warm-up'' phase.
Intuitively, this is because it takes time for comets to experience galactic tide, migrate in $q$, and arrive at $q_\crit$.
The timescale of 400 Myr is consistent with the typical galactic tide cycle period, which is on the order of $300$ Myr \citep[estimated in Equation 18 in][]{{HT1986}}.
After the first 400 Myr, we find that the simulation rate matches well with analytic predictions.
In addition, the rate stays constant with no signs of reduction over a 1 Gyr timescale.
Thus, without a companion, an Oort cloud is capable of delivering materials into the WD tidal disruption zone at a constant rate between $\dot{M}_Z \approx 5\times 10^7~\mathrm{g\cdot s^{-1}}$ and $10^8~\mathrm{g\cdot s^{-1}}$ over a 1 Gyr timescale.

Figure \ref{fig:rate_over_time_planet} shows the pollution rate over time for $\gamma=$ 2.5 and 3.5 in the case of galactic tide and a planetary mass companion. 
The simulation rate does not match as well with the analytic prediction (Equation \ref{eqn:planet_theory_rate_2}).
Recall that there are some limitations in the analytic predictions for WD pollution rate in the presence of a companion, since this framework does not include additional complicated effects induced by scattering.
This is typically off by about a factor of 2-3, consistent with what is seen in the previous subsection. 

Furthermore, we again find that the accretion rate does not significantly decline over a 1 Gyr timescale. Note that in the case where $\gamma=2.5$ (bottom panel), the accretion decreases by about a factor of 3-4 over a 1 Gyr timescale.
Note that in the context of WD pollution rate which ranges 5 orders of magnitude, this reduction factor of 3-4 over a Gyr timescale is not significant.
The reduction factor for $\gamma=3.5$ is even less, where the accretion rate only reduces by about a factor of 2.
This can be explained by the distribution of comets.
In the case of $\gamma=2.5$, there are more comets at larger at semi-major axes than $\gamma=3.5$ and thus, are easier to be ejected.
Hence, the reservoir depletes quicker at low $\gamma$ than high $\gamma$.
None of this is observed in the galactic tide only case because there, comets semi-major axes are conserved and the reservoir does not get depleted by ejection.

Finally, there are variations over time in the accretion rate over time for both cases with and without planet.
These variations are due to the limited number of comets in our simulation.
Similar to the discussion in subsection \ref{sec:galactic_tide_rate}, the variations over every Myr bin is because of our resolution-limited simulation ($\sim 10^7$ comets).
This is about 4 orders of magnitude smaller than the actual Oort cloud population of $10^{11}$ comets.
Scaling these to a full Oort cloud, the variations would be smaller by 2 orders of magnitude.
We observe that the variations per Myr bin in our simulations (Figures \ref{fig:rate_over_time}, \ref{fig:rate_over_time_planet}) are within an order of magnitude of the mean.
Thus, for a full Oort cloud, these variations would be within $10^{-1}$ of the mean rate.

In summary, we find that in the presence
of a planetary mass companion and galactic tide, an Oort cloud is capable of delivering materials at a relatively constant rate $\dot{M}_Z^\mathrm{(planet)} \approx 1-3\times 10^7~\mathrm{g\cdot s^{-1}}$ over a 1 Gyr timescale.
Depending on $\gamma$, this rate may be reduced by a factor of 3-4 over this timescale.

\section{Discussion}\label{sec:discussion}

Figure 4 of \cite{BlouinXu2022} presents observed WD pollution rates for WDs with ages ranging from 1 Gyr to 8 Gyr.
In this Figure, they show that the pollution rate ranges between $10^5$ to $10^{10}\mathrm{~g~s^{-1}}$, with the majority ranging between $10^6$ to $10^8\mathrm{~g~s^{-1}}$.
They also find that pollution rates decrease by no more than one order of magnitude over the course 8 Gyrs.
Recent observations by \cite{Mullally2024} find giant planet candidates around two polluted WDs with ages 1.5 and 5 Gyrs.
The planets have masses $1-7 M_\Jup$ and have separations $10 - 35$ AU.
Therefore, observational evidence suggests that giant planets do not significantly reduce WD pollution rate.

Through simulations, we find that an Oort cloud (with total number of objects and total mass like the Solar System) is capable of delivering materials into a WD Roche radius at a rate from $5\times 10^6$ to $10^8 \mathrm{~g~s^{-1}}$, depending on the existence of a planetary companion.
These rates are above the current detection limit at $\sim 5 \times 10^5 \mathrm{g\cdot s^{-1}}$.
Furthermore, these pollution rates can be sustained and decrease no more than a factor of 3-4 (in the existence of a planet) over a Gyr timescale.
We find that the simulated rates found here can explain a significant portion of observed pollution rates \citep[e.g.][]{BlouinXu2022} and sustain that rate over a Gyr timescale.
Our results further show that Oort cloud comets can pollute WDs in evolved systems with giant planets as observed by \cite{Mullally2024}.

In the scenario of WD-WD binary with separation $a_p = 10$ AU, the pollution rate is $\approx 3 \times 10^5\mathrm{~g~s^{-1}}$, below the detection limit.
This rate is significantly lower than other cases due to a combination of precession, scattering, and the low chance of collisions.
This rate can be used as an order-of-magnitude estimate for other stellar-mass companion cases where $M_p \sim M_*$.

We note that an advantage of the mechanism and reservoir presented here is that it is time-independent, in contrast to other works \citep[c.f.][]{Debes2012, Mustill2014, Smallwood2018}.
The pollution rate from an Oort cloud would only decrease if the reservoir is significantly depleted.
Over an 8 Gyr timescale with a comet injection rate of $\sim 10^8 \mathrm{~g~s^{-1}}$, an Oort cloud would only deplete by roughly $5\times 10^{-3} M_\oplus$ worth of materials, which is much less than the current Solar System Oort cloud reservoir of $2 M_\oplus$. 
In the presence of other perturbers like a planet or stellar flybys, the Oort cloud reservoir would be depleted faster due to ejection.
As shown previously, with a planet we find that the pollution rate can decrease at most by a factor of 3-4 in the course of 1 Gyr.

We further discuss below the scalability of the Oort cloud, the potential impacts of stellar flybys, concerns of using the Oort cloud as a reservoir for WD pollution, the effects of very close companions, wide companions, and delivery from an accretion disc into a WD.

\subsection{Robustness of the Oort cloud}

The results in this work can be scaled to other exo-Oort clouds with different $N_\mathrm{Oort}$ and $M_\mathrm{Oort}$ by:
\begin{equation}
    \Gamma_\mathrm{total}^{\mathrm{new}} = \Gamma_\mathrm{total} \cdot \left(\frac{N_\mathrm{Oort}^{\mathrm{new}}}{10^{11}}\right)
\end{equation}
\begin{equation}
    \dot{M}_Z^{\mathrm{new}} = \dot{M}_Z \cdot \left(\frac{N_\mathrm{Oort}^{\mathrm{new}}}{10^{11}}\right) \left(\frac{M_\mathrm{Oort}^{\mathrm{new}}}{2 M_\oplus}\right)^{-1}.
\end{equation}
The other parameters of an Oort cloud, $\gamma$, $a_1$ (inner edge of the Oort cloud) and $a_2$ (outer edge), do not significantly affect pollution rate. 
First, we studied various $\gamma$ in this article and find that $\gamma$ does not affect the pollution rate by more than a factor of 2 (Figures \ref{fig:tide_rate_powerlaw}, \ref{fig:tide_rate_powerlaw_planet}, \ref{fig:rate_wd_wd}).
Second, we fixed the inner semi-major axis edge at $a_1 = 3~000$ AU, based on the Solar System Oort cloud.
\cite{OConnor2023} find that varying $a_1$ changes the pollution rate by at most a factor of 4 for $10^3\mathrm{~AU} \leq a_1 \leq 10^4$ AU.
Third, we fixed the outer semi-major axis edge at $a_2 = 85~000$ AU.
This is because we have shown that this outer edge is a natural consequence of enforcing the boundary condition of the WD Hill sphere at $0.8$ pc.
Thus, $a_2$ would not be different in another Oort cloud around a typical WD.

\subsection{Stellar Flybys}

Stellar flybys are an additional mechanism which can both reduce and increase pollution rate.
First, a strong flyby (slow with small impact parameter) could potentially induce strong scattering and significantly deplete an Oort cloud.
We do not consider stellar flybys in this work, thus it is unclear to us how flybys would deplete the Oort cloud reservoir.
However, \cite{Higuchi2015} show that in simulations of Oort clouds with galactic tide and impulsive stellar flybys over different $\gamma$ structures, the $e$-folding decay timescale for the Oort cloud population is $4-18$ Gyrs.
Thus, even with stellar flybys, we still do not expect a strong (more than one order-of-magnitude) decrease in pollution rate within a Gyr timescale because the reservoir is not significantly depleted.

Second, distant flybys can stochastically perturb comets and cause them to diffuse into small pericentres.
Hence, weak flybys can act as another mechanism, in addition to galactic tide, to deliver comets from the Oort cloud into small pericentre.
This was first explored by \cite{HT1986} and used by \cite{OConnor2023} to estimate that the effects distant flybys contribute to the comet injection rate is on the order of the pollution rate from galactic tide alone.

Third, a strong flyby can cause comet showers \citep{Heisler1990}.
In the Solar System, these showers increase the injection rate of Oort cloud comets into $q \sim 10$ AU.
In the Solar System these showers have increased injection rate as much as two orders of magnitude within a few Myr -- much shorter than the galactic tidal timescale.
The precession and scattering barriers we discussed earlier cannot prevent these comet showers because comets are induced into low pericentre within one orbital period.
In the loss cones formulation, this is equivalent to a significant number of comets are in the filled loss cone regime and are able to bypass both angular momentum and scattering barriers.
In the context of exo-Oort cloud around WDs, this is perhaps another mechanism to not only deliver materials, but also potentially explain the observed spread in pollution rate (5 orders of magnitude).

\subsection{Surviving Stellar Evolution}

One major concern of having an Oort cloud as a potential material reservoir for WD pollution is that the Oort cloud might be ejected during the evolutionary process from main sequence to WDs. 
A comet at $10~000$ AU has a typical orbital speed of $\sim 0.3 \mathrm{~km~s^{-1}}$. 
This is lower than the typical speed the natal, anisotropic recoil kick a WD experiences during its rapid mass loss phase, which is about $0.75~\mathrm{km~s^{-1}}$ \citep{ElBadry2018}.
The star also undergoes mass loss at the same time. 
Thus, it is a concern if objects in the Oort cloud can be kept bounded to its central star.
\cite{OConnor2023} also investigate this question to find that a post-main-sequence evolution Oort cloud retains about 10\% of its original objects, with a fairly complex cloud structure.
Thus, an Oort cloud can remain bound, albeit with less materials, after an anisotropic mass loss during stellar evolution. 

\cite{OConnor2023} also find that the pollution rate after main-sequence evolution is reduced by about an order of magnitude for a typical kick strength of $0.75~\mathrm{km~s^{-1}}$.
Hence, if we assume a more complicated post-main-sequence evolution Oort cloud where the original Oort cloud is like our current Solar System Oort cloud, the pollution rates in our work is reduced by another order of magnitude.
This gives the post-stellar evolution pollution rate to be in the range between $10^6$ to $10^7 \mathrm{~g~s^{-1}}$, depending on the existence of a planet.
If the companion is stellar-mass, the pollution rate is much lower at $\sim 3\times 10^4 \mathrm{~g~s^{-1}}$.

However, the mass and structure of exo-Oort clouds in other main-sequence or post-main-sequence systems are unknown.
This could point to a much wider range of Oort clouds' total mass or number of objects.
In addition, Solar System Oort cloud formation and characterisation remains an active area of research.
For example, simulations of Solar System Oort cloud formation \citep[e.g.][]{Vokrouhlicky2019} include a very specific migration history of the planets following the Nice model.
The Solar System's Oort cloud total mass, number of objects, and chemical composition are heavily dependent on the early configuration of the giant planets.
Thus, an exo-Oort cloud might very well be different in number of objects, mass, and chemical composition from what we currently observe of our own Oort cloud.
Therefore, it is unclear how well we can extrapolate the mass of the post-main sequence Solar System Oort cloud to other WD planetary systems.

With the uncertainty of formation models and complicated post-main-sequence evolution, it is unclear how exo-Oort clouds look around WDs.
Therefore, in this article, we choose to simplify by only answering the question if an exo-Oort cloud with a $n(a) \propto a^{-\gamma}$ radial density profile --- like the one currently existing in the Solar System as we currently understand it --- can pollute a WD.

\subsection{Cometary Composition}

Another major concern for the Oort cloud as a potential material reservoir for WD pollution is that we mostly observe volatiles-poor polluted WD atmospheres \citep[e.g.][]{Jura2006, Jura2012, Doyle2019}, with accreting material composition resembling of asteroids or rocky planets in the Solar System.
These observed compositions are inconsistent with the volatiles-rich, icy bulk composition of typical Solar System comets.
Solar System Oort cloud comets are expected to be mostly ice because they are ejected from the protoplanetary disk beyond the ice line due to interactions with the Uranus and Neptune \citep{Vokrouhlicky2019}.
We discuss ways to reconcile these observations with our numerical predictions that Oort clouds, at least those like our own, should be able to pollute their WD.

First, the composition of objects in the Oort cloud in our Solar System and especially in other exo-Oort clouds might not be composed of only icy comets.
In our own Solar System, for example, \cite{Vida2023} recently observe a small rocky object with origins from the Oort cloud.
Simulations of the Solar System Oort cloud indicate that about 4\% of objects (up to $8\times 10^{9}$ objects) are rocky asteroids \citep{Shannon2015}.
In addition, our own Oort cloud is ``icy'' and volatile-rich because of the early configuration of the giant planets.
Because of this, it is also unclear regarding other details of Oort clouds in other systems.
We have observations of protoplanetary discs with diverse arrangements of giant planets configurations.
These giant planets do not always stay fixed beyond the ice line.
Thus, it is uncertain if objects interacting with different giant planets to be injected into exo-Oort clouds are also formed within or outside of the ice line \citep[e.g.][]{Doner2024}.
Therefore, it is problematic to assume Oort clouds around WDs or even in our own Solar System to be solely composed of volatile-rich, icy comets.

To add to the complexity, we also observe some polluted WDs with volatiles in their atmospheres \citep[e.g.][]{Farihi2013,Klein2021,Doyle2021}.
This includes a detection of a Kuiper Belt-analogue composition in a polluted WD \citep{Xu2017}.
In addition, \cite{Johnson2022} observe a polluted WD with compositions that is composed from both rocky and icy bodies.
They conclude that the unusual composition can be explained if the WD is polluted by two parent bodies, with a mix of Mercury-like composition and an icy Kuiper Belt-analogue.
In the context our discussion, if Oort clouds are composed of both rocky asteroids and icy comets as mentioned previously, they can potentially explain all these diverse composition observations.

\cite{Brouwers2023} explore in details the accretion of comets. 
They find that accretion can occur in two stages: the ices may sublimate and accrete first, before refractory minerals can reach the star.
Thus, the composition signature on a WD's atmosphere     may vary over time in a single accretion event, potentially also explain the composition diversity in polluted WDs' atmospheres.

\subsection{Accretion Disc Delivery}
We only study mechanisms to deliver comets into the Roche radius. After a comet achieves a distance $d \leq 1 R_\odot$, we assume that it is tidally disrupted and form an accretion disc around a WD \citep{Koester2014}. 
Poynting-Robertson drag can deliver materials from a accretion disc into WDs at the rate $10^{8} \mathrm{~g~s^{-1}}$ \citep{Rafikov2011}. However, we observe pollution rate up to $10^{10} \mathrm{~g~s^{-1}}$ on WDs. It is difficult to explain material delivery at such rate from a accretion disc. \cite{Okuya2023} show that the Poynting-Robertson delivery rate can be enhanced in the existence of some volatiles, to bring the delivery rate above $10^{8} \mathrm{~g~s^{-1}}$. Relating to our discussion of volatiles, it may be necessary to have some icy bodies to enhance accretion disc delivery rate to explain some observed higher pollution rates.

\subsection{Very Close Companions}

So far, we only considered companions with semi-major axes on the order of $a_p = 10$ or 100 AU.
There are observational evidence for companions at these semi-major axes \citep[e.g.][]{Veras2020, Blackman2021, Mullally2024}.
On the other hand, \cite{Vanderburg2020} present observational evidence for a $M_p \sim 1 M_\Jup$ planet orbiting its WD at $a_p \sim 4 R_\odot$.
In addition, \cite{Gaensicke2019,Veras2020b} present evidence for a planet at $a_p \sim 15 R_\odot$ orbiting a volatile-rich polluted WD.
These observations require us to analyse the scenario of a planet on a very small orbit.

First, we analyse if effects induced by this planet is important. 
For the planet found by \cite{Vanderburg2020}, we have $\zeta \approx 5 \times 10^3$, assuming typical incoming comets $a \sim 10^4$ AU.
Since $\zeta \gg 1$, torque induced by this planet does not overcome galactic tide.
In addition, since the planet is so close to the WD, the ejection loss cone does not significantly cover phase space more than the engulfment loss cone (Equation \ref{eqn:planet_theory_rate_2}).
Thus, the ejection loss cone should not significantly reduce pollution rate.
Therefore, the existence of a close-in planet with the configurations as found by \cite{Vanderburg2020} cannot reduce pollution rate of Oort cloud comets.
We expect the pollution rate to be decreased by about a factor of 2 because comets are equally likely collide into either bodies.
This is because we have two close central bodies, and the Safronov number for this planet is $\Theta \approx 0.01 \ll 1$.
Finally, the same conclusions apply to the system found by \cite{Gaensicke2019}.

We discussed the case for close stellar companions earlier in Section \ref{sec:WD-WD}, where we expect the pollution rate to be reduced by half.
Note that for WDs with close stellar companions (with separations on the order $R_\odot$ to AUs), the metal pollution on those WDs can also be explained by stellar winds from their companion \citep{Zuckerman2003, Zuckerman2014} and not necessarily by another reservoir like exo-Oort clouds.

We show using our analysis framework that close-in companions, either stellar or planetary, cannot significantly reduce Oort cloud comet delivery rate into a WD.
The analysis can be applied for other systems with varying companion masses and separations; except for when the separations are large, $a_p \geq 300$ AU.

\subsection{Widely Separated Companions}

For planets or stellar-mass companions widely separated ($a_p \geq 300$ AU), our analysis cannot be applied, at least for Oort clouds with an inner semi-major axis at $\sim 3~000$ AU like ours. As discussed earlier in various contexts, we typically assume that comets that encounter a companion have high eccentricity, $e \sim 1$ (or equivalently, $a \gg a_p$). Thus, we regularly expand expressions in this limit for simplifications. This assumption is also used in the analytic galactic tide loss cone theory.

Observationally, there is evidence for distant companions, both planetary and stellar mass, around WDs.
\cite{Luhman2011} provide observational evidence for a planetary companion with mass $M_p = 7 M_\Jup$ at a separation of $a_p \approx 2~500~$ AU.
\cite{Zuckerman2014} presents a catalogue of 17 WDs with companions separated by more than $10^3$ AU.
In addition, the WD-planet system we mentioned earlier observed by \cite{Vanderburg2020} is in fact a hierarchical triple system.
There is a distant star at $\sim 10^3$ AU forming a triple system with the WD-planet binary.

Relating these observations to WD pollution, \cite{Wilson2019} find that the occurrence rate of single polluted WDs and polluted WDs with wide stellar companions are the same.
The fact that there are polluted WDs in wide binaries is in contrast with our prediction in Section \ref{sec:WD-WD}.
We expect that wide binaries should have significantly reduced pollution rate (below detection limit) due to the difficulty of direct collisions because of a high Safronov number.
An implication of this could be that Oort clouds around wide stellar binaries are not similar to ours.
In addition, the existence of a distant massive object, potentially embedded within the Oort cloud itself, could change those exo-Oort cloud structures significantly from our own Solar System Oort cloud.

One dynamical study involving wide binaries around WDs is performed by \cite{Bonsor2015}. They propose that due to galactic tidal effects, a distant binary companion periodically becomes excited to high eccentricity bringing it closer to the WD.
During these close approaches, the stellar companion scatters other reservoirs, like an exo-asteroid belt or an exo-Kuiper Belt, into a WD and induces pollution. 

With an abundance of companions at all mass scales with orbital separations $a_p \geq 10^3$ AU and observations of polluted WDs in wide binaries, it remains interesting if and how these distant companions affect their Oort cloud and
influence the observed WD pollution rate.

\section{Conclusion}\label{sec:conclusion}

We have studied if an exo-Oort cloud can pollute a WD with galactic tide and a companion over a 1 Gyr timescale through numerical and analytic methods.
We analysed cases when there is galactic tide only, when the companion is a star, and when the companion is a planet.
We studied the dynamics of the companion, namely precession torque induced on exo-Oort cloud comets and the kick in semi-major axis they experience close-encounters with the companion.
We present a fast integration method that is capable of integrating $10^8$ comets over a long simulation time.

The conclusions presented below assume a Solar System Oort cloud, with total cloud mass $M_\mathrm{Oort} = 2 M_\oplus$ containing $N_\mathrm{Oort} = 10^{11}$ objects. 
These pollution rate results are scalable to other exo-Oort clouds with different masses and number of objects.
Our main conclusions are:

\begin{enumerate}
    \item In the absence of any companions, exo-Oort clouds like our own Solar System's can pollute WDs at a rate $\sim 5\times 10^{7} - 10^{8} \mathrm{~g~s^{-1}}$. 
    \item We find that the dimensionless quantity 
    \begin{equation}
        \zeta \sim \frac{32 \pi \sqrt{2}}{3} \cdot \frac{\rho_0}{M_\mathrm{reduced} / a_p^3} \cdot \left(\frac{q}{a_p}\right)^{3/2} \cdot \left(\frac{a}{a_p}\right)^{7/2}
    \end{equation}
    is a good indicator of the relative importance of the angular momentum change induced by a companion compared to that of galactic tide (see Figure \ref{fig:precession_efficiency}).
    When $\zeta \lesssim 1$, torque from a companion dominates galactic tide and reduces comet migration.
    When $\zeta \gg 1$, only galactic tide is important. 
    \item Stellar-mass companions with masses $M_p \geq 0.1 M_\odot$ reduce the WD pollution rate to $\sim 3\times 10^5 \mathrm{~g~s^{-1}}$ due to their strong angular momentum and scattering barriers, and a low-likelihood of direct collisions with the WD.
    \item Planetary-mass companions significantly reduce WD pollution rate as predicted by \cite{OConnor2023}.
    However, we find the simulation pollution rate to be 2-5 times higher than predicted by \cite{OConnor2023}.
    We attribute this discrepancy to some limitations in the modified loss cone theory.
    Namely, the modified loss cone theory assumes a 100\% effective ejection barrier.
    The modified loss cone also cannot account for migration in comets semi-major axes, which occurs when comets interact with the planet in the diffusive scattering regime.
    That said, we still find that the modified loss cone is a reasonable order of magnitude estimate.
    With the existence of a planet and in the absence of stellar flybys, we find that the WD pollution rate due to Oort cloud objects is reduced to about $\sim 10^{7} \mathrm{~g~s^{-1}}$, assuming a Solar System Oort cloud.
    \item The powerlaw density profile structure of the Oort cloud does not significantly affect (by more than half an order of magnitude) the pollution rate.
    \item The pollution rate without a planet stays constant over a 1 Gyr timescale. The pollution rate with a planet can decrease by a factor of $\sim 3$ over a 1 Gyr timescale.
\end{enumerate}

We discussed the advantages of the Oort cloud and the impacts of stellar flybys.
We also discussed two major concerns of using the Oort cloud as potential reservoir for WD pollution:
(i) the retention of Oort cloud objects after a strong anisotropic natal kick experienced by WDs and
(ii) observations of volatiles-poor WDs.
There are uncertainties in our current understandings of the Solar System and extrasolar Oort cloud's structure and composition.
However, these uncertainties could potentially explain the diversity in WD pollution rate and composition.
Finally, we applied our analysis framework in the context of observational evidence of close companions (orbits on the order of days) to predict their minimal effects on Oort cloud pollution, and stressed how we cannot apply this framework to widely separated companions (separation $\geq 10^3$ AU). 

We show that exo-Oort clouds can potentially pollute old WDs at observed rates over Gyr timescales, depending on the existence of a companion and Oort cloud structures.
However, one single reservoir and mechanism may not necessarily explain all instances of WD pollution \citep{Veras2024}.
With further observations of polluted WDs and characterisation of their companions \citep[e.g. using Gaia as shown in][]{Sanderson2022}, our results can be applied to constrain sources of WD pollution.

\section*{Acknowledgements}
We thank the reviewer Christopher O'Connor for helpful comments.
We thank Scott Tremaine for a careful reading of the manuscript and valueable feedback.
We thank Norm Murray, Sam Hadden, Dimitri Veras, Yanqin Wu, Garett Brown, and Michael Poon for insightful discussions.
This research has been supported by the Natural Sciences and Engineering Research Council (NSERC) Discovery Grant RGPIN-2020-04513.

\section*{Data Availability}
Data available on request.



\bibliographystyle{mnras}
\bibliography{references}



\appendix

\section{Fast Integration Method with Planet}\label{sec:direct_integration}

One problem in direct N-body integration involves lengths spanning many orders of magnitude: Typical incoming comets have semi-major axes on the order of $10^4$ AU, interacting with a companion at $10^1 - 10^2$ AU, and engulfed by the WD Roche radius at $1 R_\odot$. To accurately resolve these length scales, we would like to have the integration timestep $dt$ when the comet is far away to be large, and $dt$ to be small when the comet can interact with the companion. That is, we would like $dt$ to be adaptive in the following manners: on the order of the orbital period of the comet when far away, on the order of the orbital period of the companion when close in, and on the order of the pericentre passage timescale when close to the WD.

However, this is an issue for a numerical integrator because we always need to resolve the smallest orbit. Thus, having an $N=3$ bodies (WD-companion-comet) simulation will always cause $dt$ to be a fraction of the orbital period of the companion, not the comet. This is not ideal since a lot of time spent during the simulation will be spent on resolve the Keplerian orbit of a 2-body system.

To resolve this, we develop a method to simulate just the comet; the \texttt{REBOUND} simulation will have only $N=1$ particle,  as illustrated  in Figure \ref{fig:diagram_fast_integration}.
At each timestep, we manually set the total acceleration the comet experiences to
\begin{align}
    \ddot{x}_\mathrm{comet} &= -G M_* \left(\frac{x_*}{r_*^3}\right) -G M_p \left(\frac{x_p}{r_p^3}\right) \nonumber\\
    \ddot{y}_\mathrm{comet} &= -G M_* \left(\frac{y_*}{r_*^3}\right) -G M_p \left(\frac{y_p}{r_p^3}\right) \nonumber\\
    \ddot{z}_\mathrm{comet} &= -G M_* \left(\frac{z_*}{r_*^3}\right) -G M_p \left(\frac{z_p}{r_p^3}\right) - 4 \pi G \rho_0 z.
\end{align}
$x_p, y_p, z_p, r_p$ denote the distance between the comet and the companion. Similarly for $x_*, y_*$, etc. for the distance between comet and central star. Terms like $x, y, z, r$ can be quickly calculated analytically because the WD-companion system is a 2-body system (comet is massless). Therefore, at any time, we can calculate the total acceleration a comet experiences without requiring to simulate the central star or the companion. 

In addition, the last term in the $z$ component is from vertical galactic tide. Note that this is the full vertical galactic tidal term. This does not have the limitations of the orbit-averaged equations of motion, as mentioned earlier. Hence, we use \texttt{REBOUND} with the full vertical tide formulation to handle engulfments into the WD to properly capture all galactic tide dynamics. 

We use the adaptive timestep method \texttt{IAS15} \citep{ReinSpiegel2015} as the integrator for our $N=1$ simulation. The \texttt{IAS15} integrator is capable of adaptively changing timesteps to resolve close encounters. In addition, we use the recent improvement on \texttt{IAS15}'s adaptive method \citep{PRS2024}. We do not set a minimum timestep and allow \texttt{IAS15} to properly resolve all close-encounters at machine precision error in energy. When used with our fast integration method, the \texttt{IAS15} integrator is able to adaptively resolve timesteps as wanted.

We show some verifications of this fast direct integration method in Appendices \ref{appendix:verification1} and \ref{appendix:kuiper_belt}.

\section{Verification of the fast N-body method}\label{appendix:verification1}

\begin{figure}
    \centering
    \includegraphics[width=\the\columnwidth]{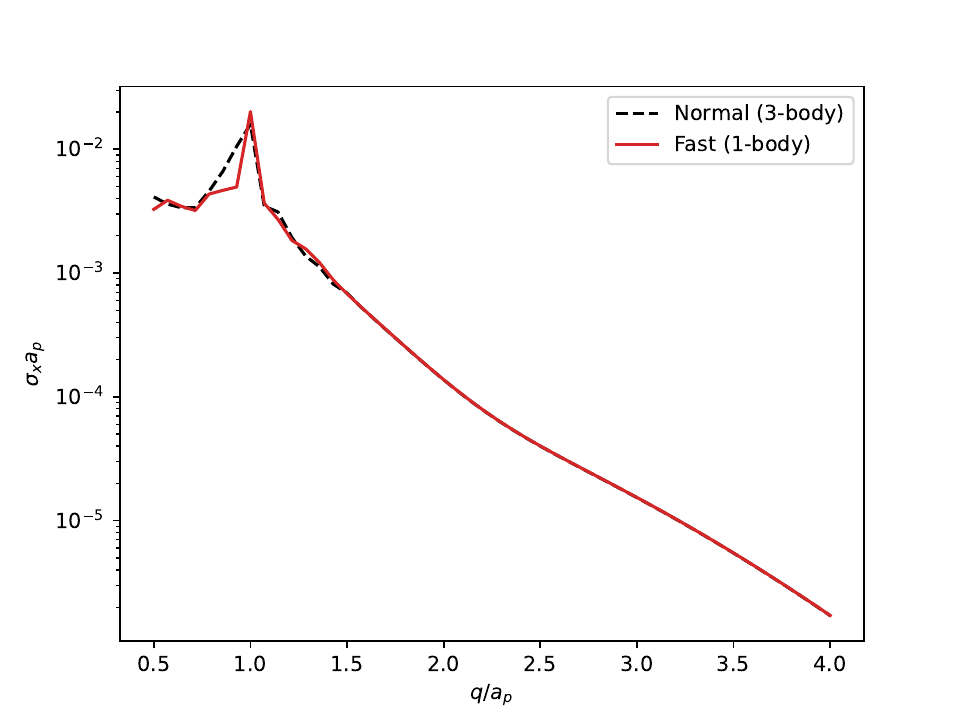}
    \caption{Verification of the fast integration method by comparing $\sigma_x$, the root-mean-square change in $x=1/a$, between two integration methods. `Normal' (black dashed line) means having 3 particles, central star, companion and comet. `Fast' (red solid line) is the integration method presented in this article with only 1 particle: a comet. In the production of this plot, the fast method is faster by about 40 times. This plot can be compared with isotropic curve Figure 9.3 in \protect\cite{Tremaine2023}.}
    \label{fig:sigmax}
\end{figure}

To verify the fast direct integration scheme (Section \ref{sec:direct_integration}) and our implementation of it in \texttt{REBOUND}, we plot Figure \ref{fig:sigmax}. 
In this Figure, we compare $\sigma_x$, the root-mean-square change in $x=1/a$, 
between our fast integration method and the full 3-body simulation. 
At each $q$, one thousand comets are initialised isotropically, with an initial semi-major axis at $a=1~000$ AU and eccentricity $e=1 - q/a$.
The planet is at $5$ AU with mass $10^{-3} M_\odot$.
Comets are integrated over one orbital period to calculate $\sigma_x$.
We can see that the fast method is able to reproduce well close encounters between comets and planet.
The results in this plot can be compared with isotropic curve in Figure 9.3 in \cite{Tremaine2023} for further verification. 
The fast method is 40 times faster.
Finally, note that this fast integration method can be applied in multiple planet systems in other applications.
We show an additional application of this with Kuiper Belt objects interactions with the outer Solar System in Appendix \ref{appendix:kuiper_belt}.

\section{An application of the fast N-body method}\label{appendix:kuiper_belt}

The fast \texttt{REBOUND} integration method can be extended to multiple planets.
One major caveat in this case is that we assume that massive objects (planets) do not interact with each other.
In other words, we assume the massive objects to be ``on rails''.
This allows us to analytically calculate the position of bodies, which is required for our fast integration method.
That being said, this method is still a good approximation when the integration timescale is less than the timescales on which planet-planet interactions begin to become important (for example, this would be the secular timescale in the Solar System).
It is possible to overcome this limitation by pre-running a simulation of the massive particles, record their positions, interpolate the positions at any $t$, and then we can use those interpolated positions for the fast integration method.

\begin{figure}
    \centering
    \includegraphics[width=\the\columnwidth]{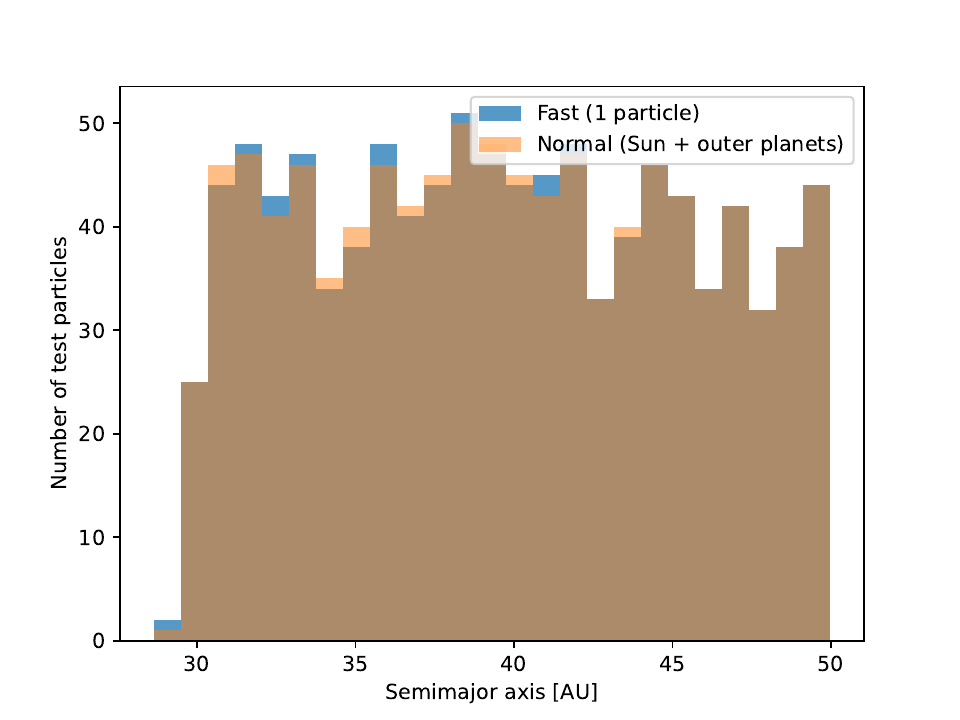}
    \caption{The Kuiper Belt semi-major axis distribution reproduced by the fast integration method (blue histogram) compared with the normal integration method (orange histogram). The simulation includes effects from the outer Solar System planets (and the Sun). The simulation is run for 100 Jupiter orbits.}
    \label{fig:kuiperbelt_semimajor_axis}
\end{figure}

In the example here, we will use the simple version (no position interpolation).
We showcase an integration of 1000 objects with semi-major axes initially distributed between $[30, 50]$ AU and inclinations drawn randomly from a Rayleigh distribution with $\langle I \rangle = 10^\circ$.
We include the Sun and the outer Solar System planets.
This integration is run for 100 Jupiter orbits.
Figure \ref{fig:kuiperbelt_semimajor_axis} shows the semi-major axis distribution of these objects.
We show here that the fast method is capable of statistically reproducing the Kuiper Belt peaks induced by resonance.

\section{Semi-major Axis Rejection Sampling}\label{appendix:rejection_sampling}

\begin{figure}
    \centering
    \includegraphics[width=1.05\columnwidth]{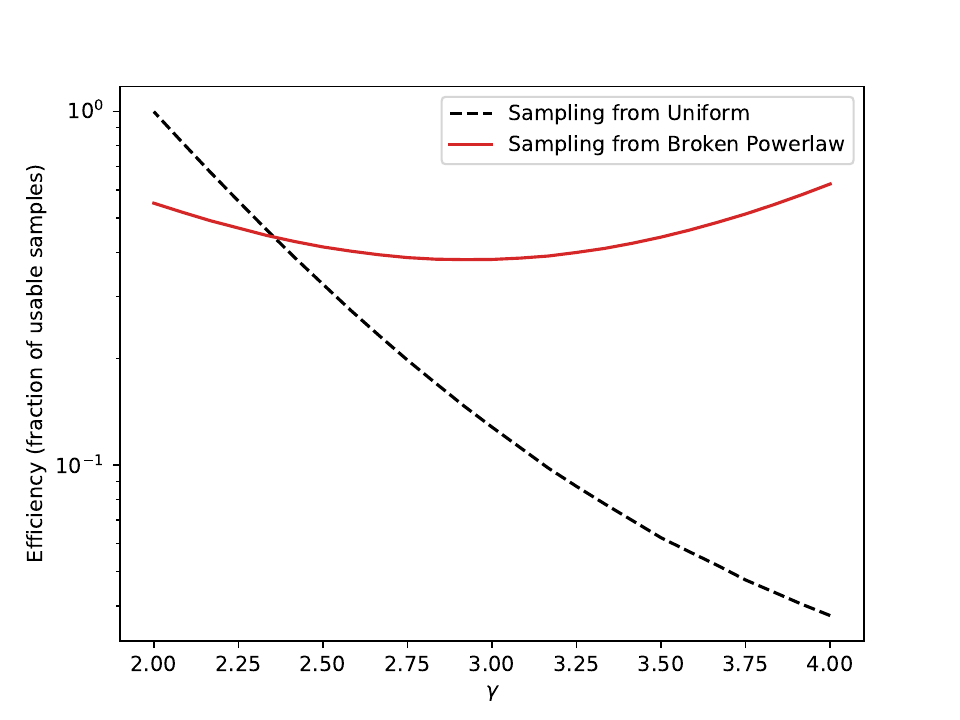}
    \caption{The efficiency of performing rejection sampling on a broken powerlaw versus rejection sampling on a uniform distribution. We can see that the using the broken powerlaw distribution to create powerlaw samples is much more effective. With the broken powerlaw rejection sampling, the number of usable samples at any particular $\gamma$ is about half of the total simulated samples.}\label{fig:rejection_sampling_efficiency}
\end{figure}

The distribution of comet semi-major axis itself follows a density profile described by a powerlaw relationship $n(a) \propto a^{-\gamma}$. We wish to study all $\gamma$ between $2 \leq \gamma \leq 4$. We could draw samples at fixed $\gamma$ and run simulation and repeat that over a grid $\gamma$. However, this increases the computation costs significantly. For example, to make Figure \ref{fig:tide_rate_powerlaw} would require running 20 simulations at different $\gamma$. To resolve this, we use a rejection sampling scheme.

We initialise comets' semi-major axes on a broken powerlaw distribution (c.f. Equation \ref{eqn:dNda}):
\begin{equation}
\dd N(a) \propto
    \begin{cases}
        a^{-\gamma_\mathrm{max}}  \cdot a^2 ~\dd a, & a_1 \leq a \leq a_\mathrm{turnover}\\
        a^{-\gamma_\mathrm{min}} \cdot a^2 ~\dd a, & a_\mathrm{turnover} \leq a \leq a_2
    \end{cases}
\end{equation}
where $\gamma_\mathrm{min} = 2$ is the minimum $\gamma$ that we can sample and $\gamma_\mathrm{max} = 4$ is the maximum. $a_\mathrm{turnover} = 17~000$ AU is the broken powerlaw turnover point. $a_\mathrm{turnover}$ is determined heuristically to maximise the efficiency of rejection sampling.

We simulate $N_\mathrm{comets}^\mathrm{sim}$ comets with semi-major axes drawn from this broken powerlaw distribution. We record their initial and final positions. Using rejection sampling, we select comets after running the simulations to construct results for any particular $\gamma$ we are interested in.

In Figure \ref{fig:rejection_sampling_efficiency}, we show the efficiency of rejection sampling from the broken powerlaw at various $\gamma$. We also compare that with rejection sampling from a uniform distribution instead. As shown, the broken powerlaw is much efficient for rejection sampling into a powerlaw distribution than a uniform distribution. We find that using the broken powerlaw as proposed gives a 50\% efficiency. That is, to construct results at any particular $\gamma$, the number of comets representative in that results is $N_\mathrm{comets}^\mathrm{eff} \sim 0.5 N_\mathrm{comets}^\mathrm{sim}$. As a consequence, we need to simulate twice the number of comets we would like to see for any particular result. However, this is still less than one simulation per $\gamma$.


\bsp	
\label{lastpage}
\end{document}
